\documentclass[pra,a4paper,10pt,twocolumn,superscriptaddress,longbibliography,nofootinbib]{revtex4-2}

\usepackage[T1,T2A]{fontenc}
\usepackage[english]{babel}

\usepackage{amssymb}
\usepackage{amsmath}
\usepackage{amsfonts}
\usepackage{graphicx}
\usepackage{bm}
\usepackage[dvipsnames]{xcolor}
\usepackage{multirow}
\usepackage{comment}
\usepackage{subfigure} 
\usepackage{color}
\usepackage{hyperref}

\DeclareMathOperator{\sgn}{sgn}

\begin{document}

\title{The Effect of Pearl Vortices on the Shape and Position of N\'eel-Type Skyrmions in Superconductor--Chiral Ferromagnet Heterostructures}

\author{S. S. Apostoloff}

\affiliation{\mbox{L. D. Landau Institute for Theoretical Physics, Semenova 1-a, 142432, Chernogolovka, Russia}}

\affiliation{Laboratory for Condensed Matter Physics, HSE University, 101000 Moscow, Russia}

\author{E. S. Andriyakhina}

\affiliation{Dahlem Center for Complex Quantum Systems and Physics Department, Freie Universit\"at Berlin, Arnimallee 14, Berlin, 14195, Germany}

\author{I. S. Burmistrov}

\affiliation{\mbox{L. D. Landau Institute for Theoretical Physics, Semenova 1-a, 142432, Chernogolovka, Russia}}

% \date{\today, v5-en}

\begin{abstract}
This review presents recent work carried out at the Landau Institute for Theoretical Physics of the Russian Academy of Sciences on the study of the effect of 
superconducting vortices on the shape and position of N\'eel-type skyrmions in superconductor--chiral ferromagnet heterostructures. Based on analytical and numerical approaches, a number of effects caused by the inhomogeneous magnetic field of the vortex have been predicted: a significant increase in the skyrmion radius, a change in its chirality in the case of 
a
coaxial configuration of the vortex and skyrmion, and 
modification of the skyrmion shape in the case of an eccentric configuration. Recent experiments studying these effects are discussed.
\end{abstract}

\maketitle

\tableofcontents

\section{Introduction\label{Sec:Intro}}

The development of modern physics is associated with a growing interest in physical objects in many-body systems that, on the one hand, represent collective excitations and, on the other hand, possess topological protection. Skyrmions, which were theoretically proposed by T. Skyrme as a model for baryons in the mid-20th century~\cite{Skyrme1,Skyrme2,Skyrme3}, are one example of such objects. Subsequently, the existence of similar topological excitations was predicted in various physical systems: two-dimensional electron gas in the quantum Hall regime~\cite{Sondhi}, spinor Bose–Einstein condensates~\cite{Stoof}, superfluid $^3$He~\cite{VolovikBook}, chiral magnetic films~\cite{Bogdanov1989}, and others. We are particularly interested in skyrmions realized in chiral ferromagnets. A detailed description of key theoretical and experimental results related to topologically nontrivial magnetic excitations (including skyrmions) in noncentrosymmetric systems can be found in the recent review~\cite{BorisovUFN2020en}.

Another well-known example of topologically protected objects in many-body systems is vortices in superconducting structures~\cite{Abrikosov-book,VolovikBook}. In %a single 
the same material, superconductivity and magnetism are typically %antagonistic 
%!%exclusive 
compete with each other %!%
and exist in different regions of the phase diagram. The creation of superconductor--ferromagnet heterostructures has made it possible to observe a number of interesting effects associated with the spatially separated coexistence of magnetism and superconductivity~\cite{Ryazanov2004,Lyuksyutov2005,Buzdin2005,Bergeret2005,Eschrig2015,Back2020,Tretiakov2021,Zlotnikov2021}. In particular, superconductor--ferromagnet heterostructures allow for the study of two topologically nontrivial objects — a skyrmion and a vortex — in a single experimental system~\cite{Petrovic2021,Machain2021,Xie2024}. Even in the absence of superconducting vortices, the presence of skyrmions can strongly influence the physics of superconductor--ferromagnet heterostructures: for example, skyrmions modify the Josephson effect~\cite{Yokoyama2015}, can induce Yu–Shiba–Rusinov-type states~\cite{Pershoguba2016,Poyhonen2016}, and affect the superconducting critical temperature~\cite{Proshin2022}.

The simultaneous presence of skyrmions and vortices in a superconductor--ferromagnet heterostructure is of particular interest because the magnetic field created by the inhomogeneous magnetization of a skyrmion 
%induces 
can induce a zero-energy quasiparticle state in the superconducting vortex core
%, which turns out to be 
being a Majorana state~\cite{Chen2015,Yang2016,Gungordu2018,Mascot2019,Rex2019,Garnier2019,Rex2020,Gungordu2022}. Such Majorana states can serve as a platform for topological quantum computing~\cite{Nothhelfer2022,Konakanchi2023}. 

There are several mechanisms for the formation of a skyrmion–vortex pair in superconductor--chiral ferromagnet heterostructures. In the case of a sufficiently good %!% 
quality %!%
interface, the proximity effect, in the presence of spin–orbit coupling, leads to the formation of a skyrmion–vortex pair~\cite{Hals2016,Baumard2019}. Regardless of the interface quality, another mechanism exists: interaction between the stray magnetic field of the vortex and the magnetization of the skyrmion~\cite{Dahir2018,Dahir2019,Menezes2019,Dahir2020}. This effect is sensitive to the spatial distribution of the skyrmion’s magnetization — i.e., whether it is of Bloch or N\'eel type. In the latter case, the stable configuration of the skyrmion–vortex pair may correspond to a finite distance between their centers~\cite{Andriyakhina2021}. This implies that when analyzing the structure of quasiparticle states in the superconducting vortex core, one cannot a priori ignore the fact that the vortex-induced magnetic field alters the spatial magnetization profile of the skyrmion (including displacing it), which in turn affects the Zeeman splitting of the quasiparticle spectrum.

Thus, a consistent microscopic analysis of Majorana states in superconductor--ferromagnet heterostructures  
%requires the ability to solve of the magnetostatic problem of
relies on an explicit form of 
the equilibrium magnetization profile in a chiral ferromagnetic film under the influence of the magnetic field created by a vortex in the superconducting film. This review presents the methods for 
%solving this problem 
solving the magnetostatic problem to find such an equilibrium profile
and recent results obtained in this context~\cite{Andriyakhina2022,Apostoloff2023,Apostoloff2024}.

Section~\ref{Sec:review} provides an overview of the problem of skyrmion–vortex interaction in superconductor--chiral ferromagnet heterostructures, including analytical methods and a numerical approach based on micromagnetic simulations. In Section~\ref{Sec:Coax}, a variational analysis of the coaxial configuration of a superconducting vortex and a N\'eel-type skyrmion is presented. Section~\ref{Sec:Eccentric} 
%presents 
%!%devotes 
is devoted %!%
to
%the approach to 
studying eccentric configurations of vortices and skyrmions and discusses various effects associated with such arrangements. Then, in Section~\ref{sec:VaV-Sk}, the developed variational approach is used to analyze the stabilization of a vortex–antivortex pair due to interaction with a skyrmion. %!% 
In %!% 
Section~\ref{sec:high_sk-v} %!%discusses 
we discuss %!% 
the influence of a vortex on higher-order skyrmions. The review concludes with a discussion of existing experimental results and final remarks.

%----------------------------------%
%------------- SECTION ------------%
\section{Skyrmion and Vortex: Perspectives and Current Status\label{Sec:review}}

%----------- SUBSECTION -----------%
\subsection{Topological Quantum Computation\label{sec:qcalc}}

The recent surge of interest in Majorana fermions in condensed matter systems has been driven by their potential use in non-Abelian quantum computation~\cite{Nayak2008}. Many researchers have focused on creating one-dimensional heterostructures in which topological $p$-wave superconductivity arises due to the proximity effect and leads to the formation of Majorana states at the boundary~\cite{Kitaev2001,Kitaev2003,Oreg2010,Lutchyn2010,Sau2010,Mourik2012,Das2012,Sun2015,Sato2017}. Another promising direction involves studying heterostructures in which a thin superconducting film is coupled to a noncollinear magnet~\cite{Nakosai2013,Chen2015}. Among the advantages of the two-dimensional realization of Majorana states in a superconductor–noncollinear magnet structure are that Majorana states can exist without precise tuning of the chemical potential, are stabilized over a much wider parameter range, and are separated by a distance determined by long-range noncollinear order, reaching macroscopic scales~\cite{Chen2015}.

The discussion of a superconductor–noncollinear magnet heterostructure in the context of Majorana states first appeared in Ref.~\cite{Nakosai2013}, where it was shown that a spiral spin structure near an $s$-wave superconductor gives rise to edge Majorana fermions with flat dispersion, while a spin configuration in the form of a skyrmion crystal generates chiral Majorana states at the sample boundary. Shortly thereafter Ref.~\cite{Chen2015} developed a more refined and unified approach that encompasses cycloidal, helical, and tilted conical orders found in multiferroics, as well as Bloch- and N\'eel-type domain walls in ferromagnetic insulators.

With the emergence of new proposals for realizing Majorana states, the need arose to study their stability and the possibility of manipulating them. Various geometries were proposed as possible implementations. Among them was a heterostructure based on a two-dimensional electron gas sandwiched between a chiral ferromagnet hosting magnetic skyrmions on one side and an $s$-wave superconductor on the other~\cite{Gungordu2018}. Another proposal considered a system in which a magnetic skyrmion with even topological charge $Q$, see Eq.~\eqref{eq:rho_Q}, was placed near an $s$-wave superconductor~\cite{Yang2016}.

However, many of the proposed geometries turned out to be difficult to implement experimentally. In particular, most known materials only support skyrmions with unit topological charge~\cite{Everschor-Sitte2018}, while the even-$Q$ states described in~\cite{Yang2016} are unstable and had not been observed until recently, when they were first found at room temperature in a multilayer structure [Co(0.2 nm)/Ni(0.7 nm)]${}_n$ with $n=4\div11$ — the number of bilayers in the heterostructure~\cite{Hassan2024}. However, Ref.~\cite{Rex2019} showed that the need to realize even $|Q| > 1$ experimentally can be elegantly avoided. If one considers a skyrmion with topological charge~$Q$ coaxial with a superconducting vortex characterized by an inhomogeneous order parameter $\Delta(\bm{r}) = e^{i b \varphi} \Delta(r)$, where ($r, \varphi$) are the radial distance and azimuthal angle from the vortex center, then it can be shown that a necessary condition for the appearance of a Majorana state is the evenness of $Q + b$. Thus, it was predicted that the most realistic case from an experimental perspective — $|Q| = |b| = 1$, corresponding to a conventional skyrmion and a single-quantum superconducting vortex — can lead to the formation of a Majorana state.

At present, such a realization of Majorana states in superconductor–chiral ferromagnet heterostructures is considered highly promising. In particular, recent works have proposed and theoretically analyzed protocols for moving Majorana states relative to each other within skyrmion–vortex pairs~\cite{Nothhelfer2022,Konakanchi2023}.

Next, we will examine in more detail the theoretical model that allows analytical investigation of the influence of a superconducting vortex on the shape and position of a skyrmion.

%----------- SUBSECTION -----------%
\subsection{Theoretical Model\label{sec:model}}

We consider a thin chiral ferromagnetic film, whose magnetic free energy is given by the expression~\cite{Bogdanov1989}:
\begin{equation}
    \frac{\mathcal{F}_{\text{magn}}[\bm{m}]}{d_F} = \int d^2 \bm{r} \bigg\{ A (\nabla \bm{m})^2 + K(1- m_z^2) + w_{\rm DM}[\bm{m}] \bigg\},
	\label{eq:MagFe}
\end{equation}
where $\bm{m}(\bm{r})$ denotes a unit vector in the direction of magnetization, $d_F$ is the film thickness, and $A > 0$ and $K > 0$ are the exchange stiffness and the effective\footnote{The contribution of the demagnetizing field is included in the effective perpendicular anisotropy constant, $K = K_0 - 2\pi M_s^2$ \protect\cite{Menezes2019,Andriyakhina2021,Kuznetsov2023PRB}.} perpendicular anisotropy constant, respectively. The density $w_{\rm DM}[\bm{m}]$ corresponds to the contribution from the relativistic antisymmetric exchange, also known as the Dzyaloshinskii–Moriya interaction (DMI). It is important to note that the form of $w_{\rm DM}[\bm{m}]$ depends on the crystalline class of the material and leads to the formation of various noncollinear structures. Here, we are interested in N\'eel-type skyrmions, which form in lattices belonging to the family of ``pyramidal'' symmetry groups $C_{nv}$. For these cases, the explicit form of the DMI contribution is:
\begin{equation}
    w_{\rm DM}
    [\bm{m}] = D \bigl [m_z \nabla \cdot \bm{m} - (\bm{m}\cdot \nabla) m_z \bigr ], \label{eq:w_cnv}
\end{equation}
where $D$ is the Dzyaloshinskii–Moriya interaction constant. Expressions for $w_{\rm DM}[\bm{m}]$ for other symmetry classes can be found in Ref.~\cite{Bogdanov1989}.

The magnetic free energy is normalized so that $\mathcal{F}_{\text{magn}}$ is zero for the ferromagnetic state with $m_z = 1$.

\subsubsection{Free Skyrmion}

In the absence of an external magnetic field, the free energy given by Eq.~\eqref{eq:MagFe} allows for the existence of metastable stationary states — skyrmions.

A single free skyrmion centered at the origin has a cylindrically symmetric magnetization profile that can be expressed in the following general form:
\begin{equation}
	\label{eq:m_free_general}
	\bm{m} = 
	[\bm{e}_{r} \cos\psi(r)  +\bm{e}_{\phi} \sin\psi(r)]\sin \theta (r)  + \bm{e}_z \cos \theta (r).
\end{equation}
Here, $\bm{e}_{r}$, $\bm{e}_{\phi}$, and $\bm{e}_{z}$ are unit vectors in the radial, azimuthal, and axial directions of the cylindrical coordinate system, where the $z$-axis is perpendicular to the interface. The angle $\psi(r)$ defines the skyrmion type: for $\psi(r) = \pi/2$ the skyrmion is of Bloch type, and for $\psi(r) = 0$ it is of N\'eel type. In what follows, we consider N\'eel-type skyrmions, as these are the ones observed in recent experiments~\cite{Petrovic2021,Machain2021,Xie2024} in configurations with a superconducting vortex. For such skyrmions, both coaxial and displaced configurations can exist~\cite{Andriyakhina2021,Andriyakhina2022,Apostoloff2024}.

Thus, the magnetization profile of a free N\'eel-type skyrmion takes the form:
\begin{equation}
	\label{eq:m_free}
	\bm{m} = 
	\bm{e}_{r} \sin \theta (r)  + \bm{e}_z \cos \theta (r),
\end{equation}
where the skyrmion angle $\theta(r)$ can be found from the Euler–Lagrange equation obtained by minimizing the free energy~$\mathcal{F}_{\text{magn}}$ with the DMI term given by Eq.~\eqref{eq:w_cnv},
\begin{eqnarray}
	\label{eq:ELE_theta_0}
	\dfrac{\ell_{w}^2}{r}\partial_r(r \partial_r\theta)-\frac{\ell_{w}^2+r^2}{2 r^2}\sin2 \theta
	+ 2\epsilon \frac{ \sin^2 \theta}{r/\ell_{w}} = 0,
\end{eqnarray}
where we introduce the dimensionless DMI parameter~$\epsilon$ and 
the length scale parameter~$\ell_{w}$, which characterizes the domain wall width:
\begin{equation}
	\epsilon = D / 2\sqrt{AK},
	\qquad
	\ell_{w} = \sqrt{A/K}.
	\label{eq:epsilon-def}
\end{equation}
Note that since $\ell_{w}$ sets the natural length scale of the problem, the ferromagnetic film can be considered thin if its thickness is small compared to this scale, $d_F \ll \ell_{w}$.

Equation~\eqref{eq:ELE_theta_0} describes a skyrmion under the following boundary conditions:
\begin{equation}
	\theta(r\to\infty)=0,
	\quad
	\theta(r=0)=\chi\pi.
	\label{eq:bound_theta0}
\end{equation} 
The first condition ensures uniform magnetization far from the skyrmion center. The second specifies that the magnetization at the center is inverted with respect to the surrounding uniform magnetization. Here, $\chi = \pm1$ denotes the skyrmion chirality. It is important to note that a free skyrmion in an isolated ferromagnetic film possesses only one chirality, determined by the sign of the DMI parameter, $\chi = \sgn(\epsilon)$. 

In this paper, we focus on the case of positive\footnote{In Section~\ref{Sec:Coax} %!%, Fig.~\ref{fig:diagram} also presents 
we present some results for $\epsilon < 0$ in Fig.~\ref{fig:diagram}.} DMI, $\epsilon > 0$. Therefore, the only stable solution to Eq.~\eqref{eq:ELE_theta_0} for a free skyrmion corresponds to chirality ${\chi = +1}$. However, in the presence of an external inhomogeneous magnetic field, skyrmions may possess either chirality — see~\cite{Apostoloff2023} and Section~\ref{Sec:Coax} for details.

A direct numerical solution of Eq.~\eqref{eq:ELE_theta_0} with boundary conditions~\eqref{eq:bound_theta0} yields the skyrmion profile. In addition, the exact solution is well approximated by the so-called 360$^{\circ}$ domain wall ansatz, ${\theta(r)\approx\theta_{R\delta}(r)}$:
\begin{equation}
	\label{eq:ansatz_free}
	\theta_{R\delta}(r)\equiv2\arctan\frac{\sinh(R/\delta)}{\sinh(r/\delta)}.
\end{equation}
Here, the parameter $R$ encodes two physical quantities: the skyrmion radius~$|R|$ and chirality~$\chi = \sgn(R)$, while $\delta$ represents the effective domain wall width forming the skyrmion boundary. The parameters $R$ and $\delta$ can be obtained by numerically minimizing the free energy $\mathcal{F}_{\text{magn}}$ using the domain wall ansatz~\eqref{eq:ansatz_free}.

\subsubsection{Ferromagnetic Film in an External Magnetic Field}

The influence of an external inhomogeneous magnetic field~$\bm{B}(\bm{r},z)$ on a thin ferromagnetic film is described by the Zeeman contribution, which is added to the free energy:
\begin{gather}
	\mathcal{F}_{Z}[\bm{m},{\bm{B}}]  = - d_F \int d^2 \bm{r} M_s \bm{m}\cdot \bm{B}
	|_{z=+0}
	,
	\label{eq:F-Sk-V-0}
\end{gather}
where $M_s$ denotes the saturation magnetization. In what follows, the external magnetic field is considered at the level of the thin ferromagnetic film, i.e., at $z = +0$. Therefore, we will omit the notation~$|_{z=+0}$ and simply write $\bm{B}(\bm{r})$, unless otherwise specified. Moreover, we assume that the magnetic field vanishes at large distances, ${\bm{B}(\bm{r}\to\infty)\to0}$, which is naturally satisfied for the stray field of one or 
%more 
several
superconducting vortices.

To determine the stable states of the ferromagnetic film in the presence of an external magnetic field $\bm{B}(\bm{r})$, one must minimize the total free energy:
\begin{equation}
   \label{eq:F_ferro}
    \mathcal{F}_{\rm ferro}[\bm{m},{\bm{B}}]=\mathcal{F}_{\text{magn}}[\bm{m}]+\mathcal{F}_{Z}[\bm{m},{\bm{B}}].
  \end{equation}

The simplest stationary state is the ``no-skyrmion'' configuration, where the magnetization of the ferromagnetic film under the influence of the inhomogeneous field~$\bm{B}$ remains nearly uniform and does not contain skyrmions or other domain boundaries. We will consider a weak field characterized by an effective strength~$\gamma$,
\begin{equation}
	\label{eq:gamma-gen-def}
	\gamma = M_s B_0 / (2K) \ll 1,
\end{equation}
where $B_0$ is a characteristic constant amplitude of the field, ${\bm{B}(\bm{r}) = B_0 \bm{b}(\bm{r})}$ in the film, and $\bm{b}(\bm{r})$ is a dimensionless coordinate-dependent %!% 
vector 
%!% 
function. Note that the definition of $B_0$ is %somewhat 
ambiguous. It should be chosen such that: (i) the effective strength~$\gamma$ in Eq.~\eqref{eq:gamma-gen-def} is estimated reasonably, and (ii) the vector~$\bm{\mu}_{\bm{b}}$ in Eq.~\eqref{eq:mu_b} is of order~$\gamma^0 = 1$. The particular choice of~$B_0$ does not affect the final result and is needed only for evaluating the effective strength~$\gamma$, since only the product $\gamma \bm{\mu}_{\bm{b}}$ enters the answer. 
In particular, it is convenient to choose $B_0$ so that the function $\bm{b}(\bm{r})$ on the characteristic length scale $\ell_w$ is of order~$\gamma^0 = 1$, which is done in Eq.~\eqref{eq:gamma-def} for the Pearl vortex.

Then, the magnetization~$\bm{m}_{\bm{b}}$ of such a stable state is approximately given by
\begin{equation}
	\label{eq:mu_b}
	\bm{m}_{\bm{b}} \approx \bm{e}_z + \gamma \bm{\mu}_{\bm{b}},
\end{equation}
where $\bm{\mu}_{\bm{b}}(\bm{r})$ is a vector of order~$\gamma^0$, orthogonal to~$\bm{e}_z$, i.e., $\bm{\mu}_{\bm{b}} \cdot \bm{e}_z = 0$.

Expanding the total free energy in Eq.~\eqref{eq:F_ferro} for~${\bm{m} = \bm{m}_{\bm{b}}}$ to second order in~$\gamma$ and minimizing it yields the Euler–Lagrange equation for~$\bm{\mu}_{\bm{b}}$:
\begin{equation}
	\label{eq:mu_b_ELE}
	\ell_w^2 \Delta \bm{\mu}_{\bm{b}} - \bm{\mu}_{\bm{b}} + \bm{b}_{||} = 0,
\end{equation}
where $\Delta$ denotes the Laplacian, and $\bm{b}_{||} = \bm{b} - \bm{e}_z b_z$ is the in-plane component of the normalized external magnetic field~$\bm{b}$.

\subsubsection{Magnetic Field Generated by a Pearl Vortex}

%%%%%%%%%%%%%%%%%%%%%%%%%%%%%%
% Figure 1
%%%%%%%%%%%%%%%%%%%%%%%%%%%%%%
\begin{figure}[!t]
	\includegraphics[width=3.5in]{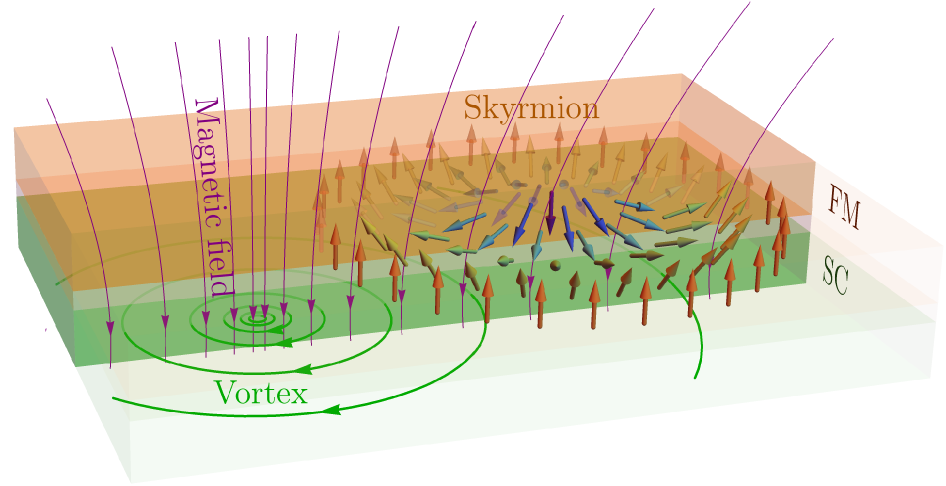}	
	\caption{Schematic illustration of a heterostructure composed of thin superconducting (SC) and ferromagnetic (FM) films, containing the corresponding topological objects — a vortex and a skyrmion — which interact via the stray magnetic field.}
	\label{fig:scheme}
\end{figure}
%%%%%%%%%%%%%%%%%%%%%%%%%%%%%%%%%

Hereafter, we assume that the magnetic field $\bm{B}(\bm{r})$ acting on the ferromagnet originates from either a single superconducting vortex or a vortex–antivortex pair. Consider a heterostructure consisting of two thin films — a superconducting and a ferromagnetic one — separated by a thin insulating layer that suppresses the proximity effect (see Fig.~\ref{fig:scheme}). We assume that the thickness $d_S$ of the superconducting film is much smaller than the London penetration depth, $d_S \ll \lambda_L$. In this case, the superconducting vortex is referred to as a Pearl vortex, and the characteristic length that defines the scale of the vortex’s stray magnetic field is $\lambda = \lambda_L^2 / d_S$, known as the Pearl length~\cite{Pearl1964}.

The magnetic field ${\bm B}_{V}$ generated above the superconducting film (at $z > 0$) by a single Pearl vortex centered at the origin is given by~\cite{Carneiro2000}:
\begin{eqnarray}
	{\bm B}_{V}  = \phi_0 \nabla 
	\int \frac{d^2\bm{q}}{(2\pi)^2} \frac{e^{-q z+i \bm{q}\bm{r}}}{q F(q)},
	\label{eq:vortex:B}
\end{eqnarray}
where $\phi_0 = hc/2e$ is the magnetic flux quantum. For arbitrary $d_S$ and $\lambda_L$, the function $F(q)$ is:
\begin{equation}
	F(q) = \lambda_L^2 \kappa \frac{(q+\kappa)^2 e^{\kappa d_S} - (q-\kappa)^2 e^{-\kappa d_S}}{(q+\kappa)e^{\kappa d_S} + (q-\kappa)e^{-\kappa d_S} - 2q},
	\label{eq:SM_Fq}
\end{equation}
with $\kappa = \sqrt{q^2 + \lambda_L^{-2}}$.

In the thin film limit $d_S \ll \lambda_L$, and not too close to the vortex center, $r \gg d_S$, the integral in Eq.~\eqref{eq:vortex:B} can be approximated using ${F(q)  %!%= 
\simeq 
%!%
1 + 2q\lambda}$. Then, the %!%Pearl vortex 
magnetic field of the Pearl vortex, %!% 
which is cylindrically symmetric about the vortex center, in the ferromagnetic film (at~${z = +0}$) takes the form:
\begin{eqnarray}
	{\bm B}_V(\bm{r}) =  
	\frac{\phi_0}{4\pi \ell_{w}\lambda}
	\left[
	b_r(r)\bm{e}_{r} + b_z(r)\bm{e}_z
	\right], \label{eq:vortex:B:thin-F}
\end{eqnarray}
where the functions $b_z(r)$ and $b_r(r)$ are defined by Eq.~\eqref{eq:vortex:B} for $r \gg d_S$:
\begin{gather}
	b_{r/z}(r) = -\ell_{w}\int\limits_{0}^{\infty} \frac{dq\,q J_{1/0}(qr)}{q + 1/(2\lambda)}.
	\label{eq:vortex:h}
\end{gather}
Here, $J_a(x)$ denotes the Bessel function of the first kind.

Note that for $r \lesssim d_S$, the magnetic field calculated using the above approximation ${F(q) = 1 + 2q\lambda}$ deviates from the more general result of Eq.~\eqref{eq:SM_Fq}. However, it turns out that $\bm{\mu}_{\bm{b}}$ is small in this region (since $\bm{\mu}_{\bm{b}} \to 0$ as $r \to 0$), and therefore, the deviation of the magnetic field near the vortex core has little effect on the behavior of the skyrmion. Thus, we will continue using the thin-film approximation even for $r \lesssim d_S$, i.e., for all $r$.

We restrict our analysis to this thin-film limit, as it allows for a number of analytical results. It is important to note that the variational method presented in Sections~\ref{Sec:Coax} and~\ref{Sec:Eccentric} is applicable to arbitrary magnetic field distributions and therefore can be used with the full expressions~\eqref{eq:vortex:B} and~\eqref{eq:SM_Fq}.

Hereafter, we assume that the characteristic sizes of the skyrmion and the distances between the centers of the skyrmion and the superconducting vortex in stable configurations are governed by the domain wall width~$\ell_w$, which is assumed to be small compared to the Pearl length~$\lambda$, achievable for sufficiently thin superconducting films:
\begin{equation}
    d_S \sim d_F \ll \ell_{w} \sim \delta \sim |R| \ll \lambda = \lambda_L^2 / d_S.
    \label{eq:assumptions}
\end{equation}
Under these assumptions, we may neglect $1/(2\lambda)$ in the denominator of the integrand in Eq.~\eqref{eq:vortex:h} and obtain simplified expressions for the vortex field:
\begin{gather}
   b_{r/z}(r) = -\ell_{w}/r, \qquad r \ll \lambda.
   \label{eq:vortex:h:approx}
\end{gather}

The characteristic magnitude of the Pearl vortex stray field, and hence the effective vortex strength~$\gamma$ defined in Eq.~\eqref{eq:gamma-gen-def}, are given by:
\begin{equation}
 \label{eq:gamma-def}
 B_0 = \dfrac{\phi_0}{4\pi \ell_{w}\lambda},
 \qquad
 \gamma = \dfrac{M_s\phi_0}{8\pi\lambda \sqrt{AK}}.
\end{equation}
In the subsequent analytical treatment, we consider $\gamma$ to be a small parameter, consistent with experimental observations~\cite{Apostoloff2023}.

Since the magnetic field of a single Pearl vortex is cylindrically symmetric, the magnetization in the ``no-skyrmion'' state from Eq.~\eqref{eq:mu_b} is also cylindrically symmetric, with the vector~$\bm{\mu}_{\bm{b}}$ aligned with~$\bm{e}_{r}$:
\begin{eqnarray}
	\label{eq:mu_theta}
	\bm{\mu}_{\bm{b}}(\bm{r}) = \theta_{\bm{b}}(r)\bm{e}_{r}.
\end{eqnarray}
Then, Eq.~\eqref{eq:mu_b_ELE} reduces to an equation for $\theta_{\bm{b}}$:
\begin{equation} \label{eq:thetagamma-ODE}
	\dfrac{\ell_{w}^2}{r}\partial_r(r \partial_r{\theta}_{\bm{b}}) - \frac{\ell_{w}^2 + r^2}{r^2} {\theta}_{\bm{b}} + {b}_{r} = 0.
\end{equation}
The solution to this equation with zero boundary conditions,
\begin{equation}
	\theta_{\bm{b}}(r=0)=0,
	\quad
	\theta_{\bm{b}}(r\to\infty)=0,
	\label{eq:bound_theta_b}
\end{equation}
can be found analytically~\cite{Apostoloff2024}:
\begin{equation}
\label{eq:theta_b:vortex:lambda}
	\theta_{\bm{b}}(r) = -2\ell_{w}\int\limits_{0}^{\infty} \frac{dq\,qJ_{1}(qr)}{(2q + 1/\lambda)[1 + (\ell_{w}q)^2]}.
\end{equation}
In the limit of large Pearl length, a simpler expression is obtained~\cite{Apostoloff2023}:
\begin{equation}
	\theta_{\bm{b}}(r) = K_1(r/\ell_{w}) - \ell_{w}/r, \qquad r \ll \lambda,
	\label{eq:theta_b:vortex}
\end{equation}
where $K_n(z)$ denotes the modified Bessel function of the second kind.

\subsubsection{High-Order Skyrmions}

To characterize different types of topological objects in a ferromagnetic film, it is common to introduce the concept of topological charge~$Q$ and its local density~$\rho(\bm{r})$. These quantities are related to the magnetization~$\bm{m}(\bm{r})$ through the following expressions:
\begin{eqnarray}\label{eq:rho_Q}
  	Q = \int d^2\bm{r}\,\rho(\bm{r}),
  	\quad	
  	\rho(\bm{r}) =  \dfrac{1}{4\pi}{\bm{m}} \cdot \left[\frac{\partial {\bm{m}}}{\partial x} \times \frac{\partial {\bm{m}}}{\partial y}\right].
\end{eqnarray}

In the previous subsection, we considered the simplest type of skyrmion with a topological charge of ${|Q|=1}$. Most theoretical and experimental studies focus on such skyrmions. However, recent numerical and physical experiments have revealed other magnetic topological structures: skyrmioniums~\cite{Seng2021}, high-order skyrmions~\cite{Hassan2024}, skyrmion bags~\cite{Yang2024}, and biskyrmions~\cite{Kathyat2024}. These structures can carry a topological charge $|Q|\neq 1$ due to various features of their spin textures.

The reasons why such nontrivial topological objects can exist in a system vary. In this review, we focus more specifically on high-order skyrmions (HOS) with $|Q| \geqslant 2$, since, as shown in Refs.~\cite{Shustin2023,Shustin2024en}, such skyrmions can be stabilized in the field of a superconducting vortex.

Radially symmetric HOS centered at the origin are described using a standard parameterization analogous to Eq.~\eqref{eq:m_free_general}:
\begin{eqnarray}
\label{eq:sk_high}
  	\bm{m} = 
  	[\bm{e}_{r}\cos\phi^{(n)} + \bm{e}_{\phi}\sin\phi^{(n)}] \sin \theta  + \bm{e}_z \cos \theta.
  	\quad
\end{eqnarray}

Here, $\phi$ and $\bm{e}_{\phi}$ are the azimuthal angle and the corresponding unit vector in polar coordinates, ${\phi^{(n)} = (n-1)(\phi - \phi_0)}$, %!%and 
where %!%
$n$ is the vorticity parameter of the skyrmion texture, related to its topological charge by $n = -Q$. It is important to note that the introduction of the azimuthal angle~$\phi_0$ effectively determines the helicity of the skyrmion.

As demonstrated in Ref.~\cite{Shustin2024en}, HOS can form stable configurations with a Pearl vortex. For more details, see subsection~\ref{sec:high_sk-v}.

\subsection{Micromagnetic Simulations\label{sec:MMM}}

\subsubsection{General Concepts}

An important tool for analyzing theoretical and experimental results is the technique of micromagnetic simulation~\cite{Donahue2007}. All simulations presented in this work were performed using the Object Oriented MicroMagnetic Framework (OOMMF) software package~\cite{OOMMF} and the Ubermag environment~\cite{Ubermag}. Below, we discuss the operating principles of numerical minimizers in these programs.

In numerical calculations, we deal with a discretized problem, where the system is placed on a lattice, and classical magnetic moments of unit length are located at the lattice nodes. A key assumption of the micromagnetic approach is that the spin direction changes only slightly between neighboring lattice nodes~\cite{Brown1963}. The expression for the system's free energy is then discretized and depends on $3N$ variables, where $N$ is the number of lattice nodes and the factor 3 corresponds to the three components of the magnetization. According to the OOMMF documentation~\cite{OOMMF}, the program's algorithm is based on the step-by-step evolution of the system's classical magnetic moments using \textit{evolvers}. Each step in the dynamic evolution or energy minimization depends on the previous (or several previous) steps, and the evolvers are responsible for updating the magnetization configuration from one step to the next. There are two types of evolvers: \textit{time evolvers}, which track the dynamics of the Landau–Lifshitz–Gilbert (LLG) equation, and \textit{minimization evolvers}, which find local minima on the energy surface using direct minimization methods.

Let us now discuss the numerical solution of the dynamic problem using the %!%Landau–Lifshitz–Gilbert 
LLG equation~\cite{LandauLifshitz1935,Gilbert2004}. In the absence of dissipation and spin-polarized current, the magnetization dynamics is governed by
\begin{equation}           
    \frac{\partial\bm{m}}{\partial t} = - |\gamma_{\rm LLG}| \bm{m} \times \bm{H}_{\rm eff}, \label{eq:LLG}
\end{equation}
where $\gamma_{\rm LLG}$ is a phenomenological constant, and $\bm{H}_{\rm eff}$ represents the effective magnetic field acting on the system. This field is defined as the variational derivative of the free energy with respect to magnetization:
\begin{equation}
    \bm{H}_{\rm eff} = - \frac{\delta \mathcal{F}[\bm{m}]}{\delta \bm{m}}. \label{eq:H_eff}
\end{equation}
Naturally, stationary solutions of Eq.~\eqref{eq:LLG} correspond to local extrema of the free energy.

Including dissipation modifies the original equation. The most well-known dissipative modifications of the lossless equation include an additional term in the Landau–Lifshitz form:
\begin{equation}           
    \frac{\partial\bm{m}}{\partial t} = - |\gamma_{\rm LL}| \bm{m} \times \bm{H}_{\rm eff} - \frac{|\gamma_{\rm LL}|\alpha}{M_s^2} \bm{m} \times (\bm{m} \times \bm{H}_{\rm eff}),
\end{equation}
where $\alpha$ is a dimensionless damping constant, typically associated with the presence of spin–orbit coupling,  
or in the Gilbert form:
\begin{equation}           
    \frac{\partial\bm{m}}{\partial t} = - |\gamma_{\rm G}| \bm{m} \times \bm{H}_{\rm eff} + \frac{\alpha}{M_s} \left(\bm{m} \times \frac{\partial\bm{m}}{\partial t} \right).
\end{equation}
These two forms are equivalent and related by the transformation ${\gamma_{\rm G} = \gamma_{\rm LL} (1 + \alpha^2)}$.

From the perspective of numerical modeling, inclusion of the dissipative term allows the system to reach equilibrium in a finite time. However, if the goal is to precisely model the dynamics of the system, the choice of $\alpha$ becomes crucial.

\subsubsection{Numerical Analysis of Skyrmion Magnetization in a Pearl Vortex Field}

Micromagnetic simulation methods are actively used to study the behavior of magnetic skyrmions in external fields. For instance, the analysis of skyrmion magnetization dynamics is of particular interest in the broader context of spin transport and in the field of skyrmionics. In this review, we focus on the behavior of skyrmions in superconductor–chiral ferromagnet heterostructures when vortices are present in the superconducting layer.

In Ref.~\cite{Menezes2019}, the possibility of manipulating skyrmions in such heterostructures using vortices was investigated. In particular, the authors numerically observed that the characteristic size of a skyrmion can increase significantly (by two orders of magnitude) in the presence of a vortex field, under certain ratios of interaction parameters, such as the strength of the DMI and the Pearl vortex core size. It is worth noting that this intriguing result did not follow from analytical approaches, and no systematic analysis was performed to determine the cause and conditions for this effect.

A numerical analysis of the stability of coaxial configurations of a skyrmion and a Pearl vortex, as well as changes in the skyrmion profile depending on system parameters, was carried out by the authors of this review in Ref.~\cite{Andriyakhina2022}. Through micromagnetic simulations and a minimal analytical model, it was found that a N\'eel-type skyrmion may be unstable with respect to a displacement of its center relative to the Pearl vortex center, and its radius may significantly increase.

\begin{figure}[!t]
	\includegraphics[width=3.5in]{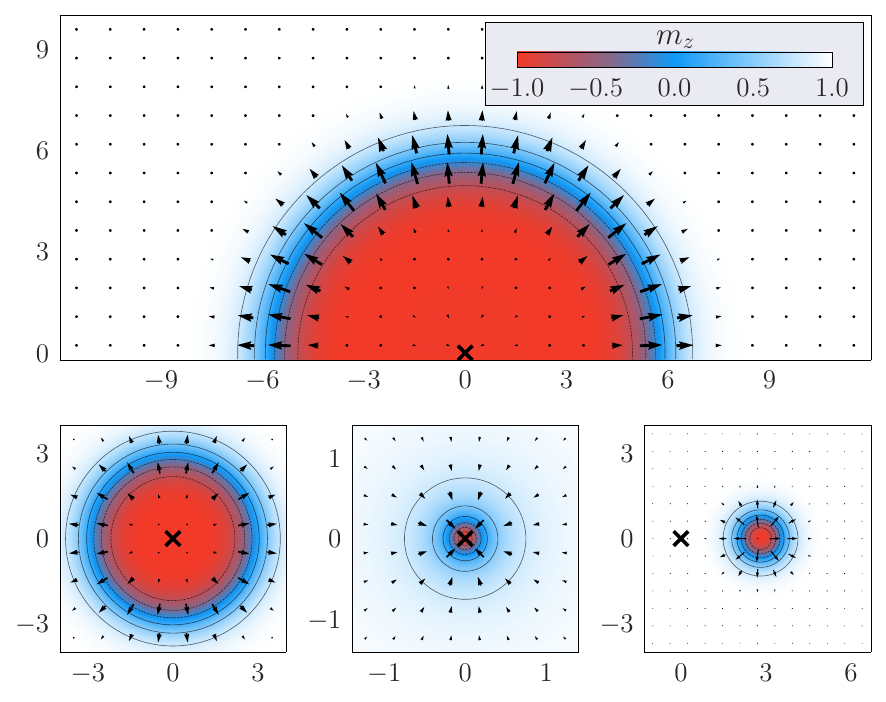}	
	\caption{Skyrmion magnetization profiles in a Pearl vortex field, obtained via micromagnetic simulation. All distributions were calculated for a DMI constant $\epsilon = 0.325$ and vortex strength $\gamma = 0.479$, confirming the existence of metastable states in the ``N\'eel skyrmion – Pearl vortex'' system. Lengths are measured in units of $\ell_w$, see Eq.~\eqref{eq:epsilon-def}. The color scale indicates the $z$-component of the total magnetization $m_z$, while the vector field shows the $x$–$y$ in-plane projection of $\bm{m}$. The vortex center is schematically marked by a cross at the origin. The top panel illustrates an abnormally large-radius skyrmion. The bottom panel shows, from left to right: a medium-size skyrmion, a small-radius skyrmion with reversed chirality, and a skyrmion displaced from the vortex center.}
	\label{fig:MMM}
\end{figure}

In our subsequent studies~\cite{Apostoloff2023,Apostoloff2024}, it was further shown that in the ``N\'eel skyrmion – Pearl vortex'' system, the magnetization can undergo chirality switching, and metastable states may exist. We will discuss this in more detail later in the review. Within this subsection, we draw attention to the micromagnetic simulation results shown in Fig.~\ref{fig:MMM}. This figure presents four metastable skyrmion states in the stray field of a Pearl vortex for fixed DMI and vortex strength parameters, $\epsilon = 0.325$ and $\gamma = 0.479$. The results illustrate the formation of three distinct coaxial configurations, as well as a configuration where the skyrmion and vortex are offset relative to each other.

In summary, numerical studies exploring skyrmions in vortex fields reveal a variety of effects associated with changes in the magnetization profile. These effects cannot be described within the framework of the standard domain wall ansatz, which motivated us to develop a new analytical approach.

%%%%%%%%%%%%%%%%%%%%%%%%%%%%%%
% Figure 2
%%%%%%%%%%%%%%%%%%%%%%%%%%%%%%
%%%%%%%%%%%%%%%%%%%%%%%%%%%%%%%%%

\subsubsection{Details of Micromagnetic Simulations}

Finally, before proceeding to the development of an analytical description of a skyrmion in a vortex field, we provide a brief overview of the micromagnetic simulations used in this work. Additional details can be found in Appendix~E of Ref.~\cite{Apostoloff2024}.

In the numerical simulations, we model the ferromagnetic film as a system of size $2L \times 2L \times dL$, composed of cells of size $dL \times dL \times dL$. All lengths are measured in units of $\ell_w$. Depending on the desired accuracy and the capabilities of the computing hardware, different values of $L$ and $dL$ were used. For the data shown in Figs.~\ref{fig:MMM} and~\ref{fig:theta_sk+v}, the following parameters were used: $L = 15$, $dL = 0.04$. For Fig.~\ref{fig:lambda-fin}, we used $L = 32$, $dL = 0.2$. Figures~\ref{fig:aRvsgamma} and~\ref{fig:phasediag} were generated using $L = 10.05$ and $dL = 0.015$. The data for Fig.~\ref{fig:aVaAV_gamma} were obtained with $L = 10$, $dL = 0.02$.

\section{Coaxial Configurations of a Skyrmion and a Vortex\label{Sec:Coax}}

We begin our study of the shape of a N\'eel-type skyrmion in the field of a Pearl vortex by considering the simplest case, where the skyrmion and the vortex are in a coaxial configuration. Such configurations were first analytically studied in Ref.~\cite{Apostoloff2023}.

\subsection{Euler-Lagrange Equation\label{sec:ELE_coax}}

In the coaxial configuration, the skyrmion exhibits a centrally symmetric magnetization; therefore, its profile can be written in the form of Eq.~\eqref{eq:m_free}. Minimizing the total free energy from Eq.~\eqref{eq:F_ferro}, we obtain the Euler-Lagrange equation (ELE) for $\theta(r)$:
\begin{gather}
\dfrac{\ell_{w}^2}{r}\partial_r(r \partial_r\theta)-\frac{\ell_{w}^2+r^2}{2 r^2}\sin2 \theta
+ 2\epsilon \frac{ \sin ^2 \theta}{r/\ell_{w}} =
\notag \\
=\gamma  (b_z\sin\theta-b_{r}\cos\theta).
\label{eq:ELE_theta_coax}
\end{gather}

\begin{figure}[t]
\centerline{\includegraphics[width=0.5\textwidth]{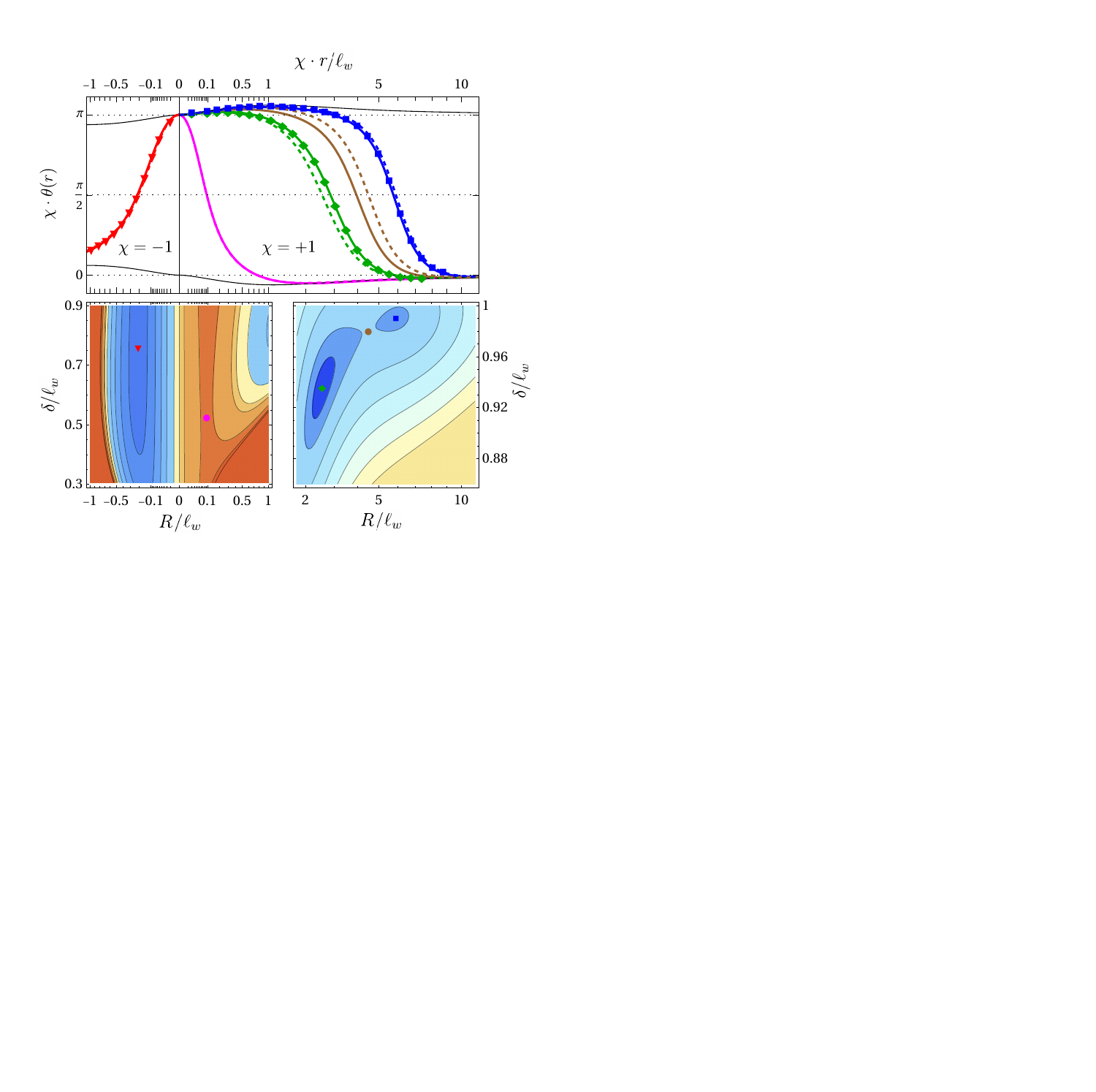}}
\caption{\textit{Top panel.} Skyrmion angle $\theta(r)$ as a function of radial coordinate~$r/\ell_w$, with sign corrected by multiplication by the skyrmion chirality $\chi=\pm1$, calculated for parameters $\epsilon=0.325$ and $\gamma=0.479$. Solid and dashed curves represent results obtained via exact solution of the ELE~\eqref{eq:ELE_theta_coax} and using the variational method with the ansatz of the form~\eqref{eq:coax-ansatz}, respectively; symbols (squares, triangles, and diamonds) are extracted from micromagnetic simulations, see Fig.~\ref{fig:MMM}. The horizontal axis uses a quadratic scale. \textit{Bottom panels:} Free energy~\eqref{eq:MagFe} of the ferromagnet with a skyrmion whose magnetization is described by the ansatz~\eqref{eq:coax-ansatz}, shown as a function of the parameters~$R/\ell_w$ and $\delta/\ell_w$. Minima are indicated by the same symbols as in the top panel, while saddle points are shown with circles. Color gradient is in arbitrary units. }
\label{fig:theta_sk+v}
\end{figure}

Equation~\eqref{eq:ELE_theta_coax}, supplemented by the boundary conditions~\eqref{eq:bound_theta0}, can be solved numerically. Note that unlike the case of a free skyrmion, where the solution is unique and possesses a definite chirality $\chi=\sgn(\epsilon)$, the presence of an external inhomogeneous magnetic field allows for multiple solutions to the ELE, including both chiralities $\chi=\pm1$. Moreover, some of these solutions may correspond not to minima of the free energy~\eqref{eq:F_ferro}, but to saddle points, see Fig.~\ref{fig:theta_sk+v}. In the latter case, the ELE solution does not describe a stable skyrmion state and should be discarded after proper verification.

%%%%%%%%%%%%%%%%%%%%%%%%%%%%%%
% Figure 3
%%%%%%%%%%%%%%%%%%%%%%%%%%%%%%
%%%%%%%%%%%%%%%%%%%%%%%%%%%%%%%%%%%

These features lead to additional technical challenges in numerically determining the stable states. Therefore, it is convenient to use a variational approach, in which the skyrmion profile is given by a fixed function containing several parameters that are determined by 
%minimizing 
the minimization of
the free energy.

\subsection{Variational Approach\label{sec:ansatz_coax}}

As shown in Ref.~\cite{Apostoloff2023}, the exact solution of Eq.~\eqref{eq:ELE_theta_coax} with boundary conditions~\eqref{eq:bound_theta0} can be effectively approximated by the following expression (hereafter referred to as the coaxial ansatz):
\begin{equation}
	\theta(r)\approx\theta_{R\delta}^{\gamma}(r)\equiv
	\theta_{R\delta}(r)+\gamma{\theta}_{\bm{b}}(r)\cos\theta_{R\delta}(r),
	\label{eq:coax-ansatz}
\end{equation}
where $\theta_{R\delta}$ is the domain wall ansatz given in Eq.~\eqref{eq:ansatz_free}, and $\gamma$ and ${\theta}_{\bm{b}}$ are defined by Eqs.~\eqref{eq:gamma-def} and~\eqref{eq:theta_b:vortex}, respectively, and arise due to the presence of the superconducting vortex.

It is useful to rewrite the ansatz~\eqref{eq:coax-ansatz} in terms of the magnetization. Indeed, if we add the vector $\gamma\bm{\mu}_{\bm{b}}$---which describes the deviation of the homogeneous magnetization under the influence of the magnetic field~$\bm{B}$ [see Eqs.~\eqref{eq:mu_b} and~\eqref{eq:mu_theta}]---to the magnetization vector~$\bm{m}_{R\delta}$ described by the domain wall ansatz~\eqref{eq:ansatz_free}, and then normalize the sum to unit length, the resulting unit vector (to linear order in $\gamma\ll1$) corresponds to a skyrmion with profile $\theta_{R\delta}^{\gamma}(r)$ from Eq.~\eqref{eq:coax-ansatz}:
\begin{equation}
	\dfrac{\bm{m}_{R\delta}+\gamma \bm{\mu}_{\bm{b}}}{|\bm{m}_{R\delta}+\gamma \bm{\mu}_{\bm{b}}|}\approx
	\bm{e}_{r} \sin \theta_{R\delta}^{\gamma}  + \bm{e}_z \cos \theta_{R\delta}^{\gamma}.
	\label{eq:m+mu:coax}
\end{equation}

The qualitative idea of the ansatz~\eqref{eq:coax-ansatz} is as follows. It is natural to expect that the skyrmion magnetization is, to leading order, described by the 360\textdegree~domain wall ansatz. However, due to the relatively weak magnetic field ($\gamma \ll 1$), the magnetization at each point is additionally rotated. This means one can write ${\theta(r)\approx\theta_{R\delta}(r)+\delta\theta(r)}$. To determine the rotation angle~$\delta\theta(r)$, we consider the magnetization in three regions: near the skyrmion center, at $r\sim |R|$, and far from the origin. Near the center or far away, the magnetization is nearly homogeneous ($m_z \approx \pm 1$), and its change is mainly governed by the vortex field. Thus, ${\delta\theta(r)\approx \pm\gamma\theta_{\bm{b}}(r)}$, directly following from the solution to Eq.~\eqref{eq:thetagamma-ODE} for the “no-skyrmion” state. In the intermediate region, ${m_z\approx\cos\theta_{R\delta}(r)}$, so the rotation angle is naturally smoothed, yielding ${\delta\theta(r)\approx \gamma\theta_{\bm{b}}(r)\cos\theta_{R\delta}(r)}$, which gives the ansatz~\eqref{eq:coax-ansatz}.

It is important to note that the rotation angle~$\delta\theta$ remains small for $\gamma\ll1$, which 
% may suggest 
makes it seems
that the ansatz in Eq.~\eqref{eq:coax-ansatz} only describes a slight deviation from a free skyrmion. Nevertheless, it turns out that for not-too-small~$\gamma$, significant changes in the skyrmion radius~$R$ can occur, and even solutions with opposite chirality may emerge~\cite{Apostoloff2023}. Indeed, the coaxial ansatz~\eqref{eq:coax-ansatz} should be substituted into the free energy expression~\eqref{eq:F_ferro} and minimized with respect to the two parameters~$R$ and~$\delta$, without assuming proximity to the free skyrmion parameters. This variational method proves 
to be
faster and more efficient than direct numerical solution of Eq.~\eqref{eq:ELE_theta_coax}. Moreover, it yields reliable results over a wide range of parameters~$\epsilon$ and~$\gamma$~\cite{Apostoloff2023}.

The main panel of Fig.~\ref{fig:theta_sk+v} shows results from the exact ELE solution (solid curves), the variational approach (dashed curves), and micromagnetic simulations for $\epsilon=0.325$ and $\gamma=0.479$. Thin solid %!%curves 
curves show $\theta(r)=\gamma\theta_{\bm{b}}(r)$ and $\theta(r)=\pi-\gamma\theta_{\bm{b}}(r)$, closely matching skyrmion profiles at small and large~$r$. The three approaches are in good agreement.

As seen from the inset in Fig.~\ref{fig:theta_sk+v}, the normalized free energy as a function of $R$ and $\delta$ may exhibit multiple minima: two minima with positive chirality, $\chi=+1$ (green diamond and blue square), and one minimum with negative chirality, $\chi=-1$ (red triangle). These minima may correspond to stable skyrmion states. Notably, the radius of the stable skyrmion (blue square in Fig.~\ref{fig:theta_sk+v}) is $R\approx5.7\ell_w$, about 14 times larger than the radius of the skyrmion $R_0\approx0.41\ell_w$ for $\gamma=0$. In addition, the free energy has two saddle points (brown and purple circles), corresponding to curves~1 and~3 in the main panel and representing unstable skyrmion states that cannot be stabilized in micromagnetic simulations.

\subsection{Analysis of the results\label{sec:results_coax}}

\begin{figure}[!t]
	\includegraphics[width=3.5in]{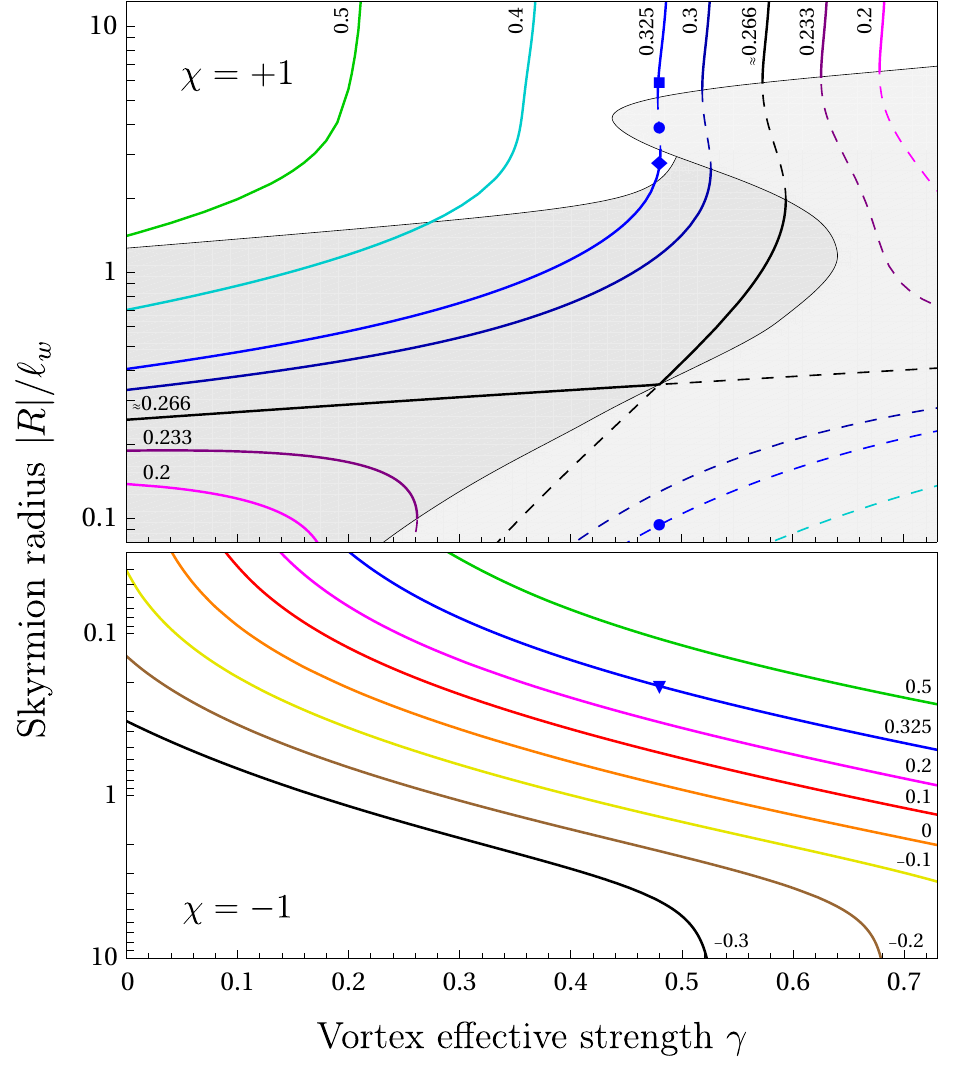}	
	\caption{Dependence of the skyrmion radius~$|R|/\ell_w$ on the effective vortex strength~$\gamma$ for several values of the DMI parameter $\epsilon$ in a semi-logarithmic scale, shown for chiralities~$\chi=\pm1$. Solid and dashed lines correspond to minima and saddle points of $\mathcal{F}[R,\delta]$, respectively. The region of unstable coaxial configurations and the region of saddle points are shaded in dark and light gray, respectively. The values of $\epsilon$ used in the figure are labeled next to the curves. Symbols (square, diamond, and triangle) correspond to stable skyrmions from Fig.~\protect\ref{fig:theta_sk+v}, and circles indicate saddle-point solutions.}
	\label{fig:diagram}
\end{figure}

Since the choice of parameters $\epsilon$ and $\gamma$ corresponds to a specific heterostructure, variations in these parameters can lead to a different number of extrema in the free energy and, consequently, to a different number of possible stable skyrmion--vortex configurations. To explore this, Fig.~\ref{fig:diagram} shows the dependence of $R$ on $\gamma$ for several values of $\epsilon$. Solid and dashed curves in the $(\gamma,R)$ plane correspond to minima and saddle points of the free energy~\eqref{eq:F_ferro} as a function of $R$ and $\delta$, respectively. The region of unstable saddle-point configurations is marked with light gray shading.

Note that even those solutions of the Euler--Lagrange equation~\eqref{eq:ELE_theta_coax} that correspond to a minimum of the free energy as a function of $R$ and $\delta$ do not necessarily represent stable skyrmion--vortex configurations. This is because the derivation of Eq.~\eqref{eq:ELE_theta_coax} assumes that the center of the skyrmion coincides with the center of the Pearl vortex, which may not correspond to a stable state for certain $\epsilon$ and $\gamma$ values, since the skyrmion center may shift away from the vortex core. As shown in Refs.~\cite{Andriyakhina2021,Andriyakhina2022,Apostoloff2024}, a skyrmion--vortex pair can remain stable when the skyrmion is located at a finite distance~$a$ from the Pearl vortex. These displaced configurations are discussed in more detail in Sec.~\ref{Sec:Eccentric}. For now, we state that coaxial configurations with chirality $\chi=+1$ are unstable for $\epsilon<\epsilon_{\rm cr}\approx 0.49$ and $\gamma<\gamma_{\rm cr}^-(\epsilon)$, see Fig.~\ref{fig:phasediag}. In this case, the skyrmion is repelled by the vortex. If either of these conditions is not met, the coaxial configuration is stable with respect to displacement. In Fig.~\ref{fig:diagram}, unstable configurations are shaded dark gray, while stable coaxial configurations are shown on a white background.

The diagram in Fig.~\ref{fig:diagram} exhibits several interesting features. First, all curves for skyrmions with chirality~${\chi=+1}$ (top panel) lie in the quadrants formed on the $(\gamma,R)$ plane by the curve with $\epsilon=\epsilon_{\rm sep}\approx 0.266$ (black line). For $\epsilon<\epsilon_{\rm sep}$, the $R(\gamma)$ curves lie in the bottom-left and top-right quadrants, while for $\epsilon>\epsilon_{\rm sep}$, they are located in the top-left and bottom-right quadrants.

Second, for $0.3\lesssim \epsilon\lesssim 0.35$, there are values of $\gamma$ for which two skyrmions with chirality $\chi=+1$ can coexist. An example of such a situation is shown in Figs.~\ref{fig:MMM} and~\ref{fig:theta_sk+v}, where the inset of the latter shows two free energy minima corresponding to positive chirality.

Third, for every pair of $\epsilon$ and $\gamma$, there exists a skyrmion with chirality ${\chi=-1}$ and a certain radius~$|R|$, see the bottom panel of Fig.~\ref{fig:diagram}. However, for small $\gamma$ and $\epsilon>0$, the skyrmion radius becomes extremely small ($R\lll\ell_{w}$). In this case, the expression for the free energy~\eqref{eq:MagFe} becomes invalid, and such small-radius solutions are excluded from Fig.~\ref{fig:diagram}. As can be seen in the figure, the radius~$|R|$ for skyrmions with~${\chi=-1}$ increases monotonically with~$\gamma$.

Finally, for each $\epsilon$ and both chiralities $\chi=\pm1$, there exists a critical value $\gamma_{\infty}^\pm(\epsilon)$. As $\gamma$ increases and approaches $\gamma_{\infty}^\pm$, the skyrmion radius grows substantially. For $|R|\gg\delta\approx\ell_{w}$, the free energy can be approximated as
\begin{equation}
	\label{eq:gamma_pm}
 \dfrac{\mathcal{F}_{\rm ferro}[R]}{8\pi A d_F}\approx
 (1\mp\epsilon\pi/2)\frac{|R|}{\ell_{w}} +\gamma \ell_{w}^{-2}\int\limits_0^{|R|}d r\, r b_z(r).
\end{equation}
The first term corresponds to the domain wall energy separating the inner and outer regions of the skyrmion, while the second term arises from the energy of the skyrmion’s inner region, where the magnetization ${m_z\approx -1}$ is opposite to the main ferromagnetic state, ${m_z\approx +1}$. For ${\ell_{w}\ll|R|\ll\lambda}$, we approximate $b_z(r)\approx -\ell_w/r$, see Eq.~\eqref{eq:vortex:h:approx}, and estimate the second term in Eq.~\eqref{eq:gamma_pm} as $\gamma|R|/\ell_{w}$. Thus, the critical value~$\gamma$ can be approximated as
\begin{equation}
	\gamma_{\infty}^\pm(\epsilon)\approx 1\mp\epsilon\pi/2>0.
    \label{eq:gamma_inf}
\end{equation}
If $\gamma \gtrsim\gamma_{\infty}^\pm$, the skyrmion radius becomes comparable to or exceeds the Pearl length, ${|R|\gtrsim\lambda\gg\ell_{w}}$, and the minimum of the free energy is determined by the condition ${-b_z(|R|)|R|/\ell_{w}\approx \gamma_{\infty}^\pm/\gamma}$, which requires taking into account the Pearl vortex field dependence on $\lambda$. A more detailed analysis of such configurations is presented in the next subsection.

\subsection{Skyrmion radius at finite Pearl length}

In the previous subsections, we assumed that the Pearl length is much larger than the skyrmion radius, i.e., $\lambda \gg R$. This assumption is justified when the parameters~$\epsilon$ and $\gamma$, which describe the Dzyaloshinskii--Moriya interaction and the influence of the superconducting vortex, are relatively small. As shown in Fig.~\ref{fig:diagram} and discussed at the end of the previous subsection, increasing these parameters leads to a growth in skyrmion size, eventually making it comparable to $\lambda$. Furthermore, from the perspective of applying theoretical results to real experiments, one should consider that $\lambda$ is related to~$\gamma$ via:
\begin{equation}
    \zeta \equiv \dfrac{\lambda\gamma}{\ell_w} = \dfrac{M_s \phi_0}{8\pi A}, \label{eq:zeta}
\end{equation}
where the introduced parameter $\zeta$ depends only on the material parameters of the ferromagnetic film and is typically of the order of $10$ (see Table~\ref{Tab:exp}). Therefore, in a certain parameter range, the finite value of $\lambda$ must be taken into account.

It is known~\cite{Tanguy2001} that the magnetic field of the vortex from Eq.~\eqref{eq:vortex:h} can be well approximated by:
\begin{equation}
	b_r(r)\approx -\dfrac{\ell_{w}}{r[1+r/(2\lambda)]}, \quad
	b_z(r)\approx -\dfrac{\ell_{w}}{r[1+r/(2\lambda)]^2}.
    \label{eq:vortex:h:simple}
\end{equation}
This allows us to write a fairly simple expression for $\theta_{\bm{b}}$, analogous to Eq.~\eqref{eq:theta_b:vortex}:
\begin{equation}
	\theta_{\bm{b}}(r) = K_1(r/\ell_{w}) - \dfrac{\ell_{w}}{r[1 + r/(2\lambda)]} - \frac{\exp(-r/\ell_{w})}{2 \lambda},
	\label{eq:theta_b:vortex:simple}
\end{equation}

By using Eqs.~\eqref{eq:vortex:h:simple} and~\eqref{eq:theta_b:vortex:simple} in the variational approach outlined in Sec.~\ref{sec:ansatz_coax}, we can determine a more accurate dependence of the skyrmion radius~$R$ on the effective vortex strength~$\gamma$. Furthermore, to analyze this dependence for large radius, we can use Eq.~\eqref{eq:gamma_pm}. 
As already mentioned, the free energy minimum is then determined by the condition ${-b_z(|R|)|R|/\ell_{w} \approx \gamma_{\infty}^\pm/\gamma}$. Using Eq.~\eqref{eq:vortex:h:simple}, we find that the minimum of $\mathcal{F}_{\rm ferro}$ is reached at a skyrmion radius
\begin{equation}
	|R| \approx 2\lambda \left(\sqrt{\gamma/\gamma_{\infty}^\pm} - 1\right)
    = 2\zeta\ell_w \left(\dfrac{1}{\sqrt{\gamma\gamma_{\infty}^\pm}} - \dfrac{1}{\gamma} \right). 
    \label{eq:R_gamma}
\end{equation}
Note that this expression remains valid for all $|R| \gg \ell_w$, which is fulfilled when $\gamma \gtrsim \gamma_{\infty}^\pm$.

\begin{figure}[!t]
	\includegraphics[width=\columnwidth]{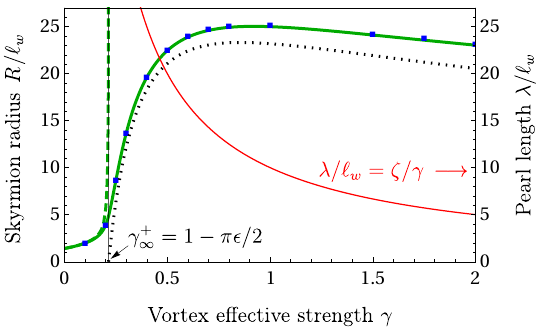}	
	\caption{Dependence of the skyrmion radius $R/\ell_w$ with positive chirality on the effective vortex strength~$\gamma$ for $\epsilon=0.5$, calculated in the limits of infinite (green dashed line) and finite (green solid line) Pearl length, as well as its asymptotic form~\eqref{eq:R_gamma} for $R\gg\lambda$ (black dotted line). Results of micromagnetic modeling are shown as blue squares. The dependence of Pearl length $\lambda = \zeta/\gamma$ on $\gamma$ is shown as a thin red line.}
	\label{fig:lambda-fin}
\end{figure}

Figure~\ref{fig:lambda-fin} shows the dependence of the skyrmion radius $R/\ell_w$ of positive chirality for $\epsilon=0.5$ on the effective vortex strength~$\gamma$, comparing the cases of infinite and finite Pearl length with the micromagnetic modeling results and asymptotic formula~\eqref{eq:R_gamma}.

We also highlight an interesting feature of the $|R|$ versus $\gamma$ dependence given in Eq.~\eqref{eq:R_gamma}: it is non-monotonic and reaches a maximum at $\gamma = 4\gamma_{\infty}^\pm$:
\begin{equation}
    |R_{\rm max}^{\pm}| = \dfrac{\zeta\ell_w}{2\gamma_{\infty}^\pm} = \dfrac{\zeta\ell_w}{2 \mp \epsilon\pi},
\end{equation}
which is determined solely by the ferromagnetic parameters and not by the superconductor. In other words, there exists an upper limit on the skyrmion size that can be achieved in a given ferromagnetic film due to the influence of a Pearl vortex.

\section{Eccentric configurations\label{Sec:Eccentric}}

In this section, we study the state of an isolated N\'eel-type skyrmion placed in an external inhomogeneous magnetic field. An analytical investigation of such states is significantly complicated by the challenge of choosing an appropriate magnetization profile~$\bm{m}$. The exact approach requires deriving the Euler–Lagrange equations and setting boundary conditions, $\bm{m} = -\bm{e}_z$ at the skyrmion center and $\bm{m} = \bm{e}_z$ far from the center. However, solving such equations, which are vector partial differential equations, numerically requires computational resources comparable to those needed for the micromagnetic simulations.

The simplest approach to describe an eccentric configuration is to compute the Zeeman contribution to the energy, which depends on the distance between the skyrmion and vortex centers, while neglecting the deformation of the skyrmion due to the vortex. This approach corresponds to the theoretical framework developed in this review in the formal limit $\gamma \to 0$. It was precisely this method that was employed in Ref.~\cite{Andriyakhina2021}, where it was found that a N\'eel-type skyrmion may energetically prefer to position itself at a finite distance from the vortex center. Qualitatively, this behavior arises due to the following physical reasoning. The nonuniform skyrmion magnetization induces a supercurrent in the superconducting film. This current has only an azimuthal component, which can vanish at a certain distance from the skyrmion center. Since the supercurrent acts on the superconducting vortex via the Lorentz force, the vortex can be in equilibrium only at points where the supercurrent vanishes. It can be shown~\cite{Abrikosov-book} that the supercurrent magnitude is determined by the derivative of the Zeeman energy with respect to the skyrmion-vortex distance. Therefore, a zero of the supercurrent corresponds to a minimum in the Zeeman energy. As we will see below, the approximation $\gamma \to 0$ used in Ref.~\cite{Andriyakhina2021} proves
% too 
to be quite 
crude, and a more refined theory is required for an accurate description of eccentric configurations.

As an alternative to an exact numerical solution, we employ a variational approach similar to that described in Sec.~\ref{sec:ansatz_coax}, based on a sufficiently simple analytical ansatz that accurately approximates exact Euler–Lagrange solutions or micromagnetic simulation results. Following Ref.~\cite{Apostoloff2024}, we will introduce, explain, and justify the choice of such an ansatz for the skyrmion magnetization under a weak, spatially varying magnetic field~$\bm{B}(\bm{r})$ in the general case, and then apply this variational method to describe eccentric skyrmion-vortex configurations. Furthermore, based on the obtained results, Sec.~\ref{sec:VaV-Sk} will analyze the stability of a vortex–antivortex pair due to interaction with a skyrmion.

\subsection{Variational approach\label{sec:var_appr}}

In this subsection, we describe the idea behind constructing an ansatz for the magnetization of a skyrmion subjected to a weak, spatially inhomogeneous external magnetic field. This ansatz is defined by three parameters: the radius~$R$, the effective domain wall width~$\delta$, and the position~$\bm{a}$ of the skyrmion relative to the magnetic field. To justify this construction, we expand the total free energy~\eqref{eq:F_ferro} to second order in the small effective field strength~$\gamma$ and analytically minimize it. As a result, we derive the expressions for the ansatz, see Eqs.~\eqref{eq:ansatz_eccentric1}, \eqref{eq:bar-m}, and~\eqref{eq:tilde-m}, and simplify the full free energy to the form of one-dimensional integrals, see Eqs.~\eqref{eq:ferro_expand2}, \eqref{eq:ferro_expand1_explicit}, and~\eqref{eq:F2-final}. These simplified expressions can then be used for the numerical determination of the parameters~$R$, $\delta$, and~$\bm{a}$ that describe the skyrmion magnetization.

It is also useful to outline  
%the overall structure 
the key stages
of the calculation. Firstly, in order to determine the parameters~$R$ and~$\delta$ at a fixed skyrmion position~$\bm{a}$, it is sufficient to expand the free energy only to first order in~${\gamma \ll 1}$, as shown in Eq.~\eqref{eq:ferro_expand1_explicit}, and then minimize it. At this stage, one may neglect skyrmion deformation and assume that its magnetization remains radially symmetric. In this case, we obtain the skyrmion radius and domain wall width as functions of the position, $R(\bm{a})$ and $\delta(\bm{a})$, since the magnetic field is inhomogeneous.

Secondly, in order to determine~$\bm{a}$, second-order contributions in the free energy expansion must be taken into account, see Eqs.~\eqref{eq:ferro_expand2} and~\eqref{eq:F2-final}. These contributions capture the deformation of the skyrmion's shape. The total free energy must then be minimized with respect to~$\bm{a}$, using~$R$ and~$\delta$ as functions of the skyrmion position.

Note that the stages described above imply that the minimization procedure can be performed in two different ways. The direct method involves searching for the minimum simultaneously with respect to all three parameters: $R$, $\delta$, and~$\bm{a}$, which yields the stable skyrmion configuration with minimal energy, similar to micromagnetic simulations.

The second method involves a two-step procedure. First, the free energy is minimized only with respect to~$R$ and~$\delta$ at fixed~$\bm{a}$. Then, the resulting function of~$\bm{a}$ is analyzed to identify local minima corresponding to metastable skyrmion positions. This approach differs from micromagnetic modeling 
because
%in that 
it facilitates the identification of multiple potentially metastable extrema.

A detailed comparison of the analytical results and micromagnetic simulations will be presented in Sec.~\ref{sec:results_ecc}.

\subsubsection{Formulation of the magnetization ansatz\label{sec:ansatz_ecc}}

For convenience in the subsequent derivation, we shift the origin of coordinates to the center of the skyrmion, denoted by the point~$\bm{a}$. As a result, the shifted external magnetic field depends on~$\bm{a}$ as a parameter, ${\bm{B}^{\bm{a}}(\bm{r})=\bm{B}(\bm{r}_{\bm{a}})}$, where ${\bm{r}_{\bm{a}}=\bm{r}+\bm{a}}$.

The central idea of the ansatz construction is to find a leading-order approximation for the skyrmion magnetization~$\bm{m}$ in the form of a radially symmetric function~$\bar{\bm{m}}$, cf. Eq.~\eqref{eq:m_free}:
\begin{equation}
	\bar{\bm{m}} = \bm{e}_{r} \sin \bar{\theta}(r) + \bm{e}_z \cos \bar{\theta}(r),
	\label{eq:bar-m}
\end{equation}
where $\bar{\theta}(r)$ is the skyrmion angle.

As will be shown in Sec.~\ref{sec:F1}, the angle~$\bar{\theta}$ can be interpreted as the skyrmion angle in the effective magnetic field~$\bar{\bm{B}}^{\bm{a}}$, i.e., the angular average of ${\bm{B}}^{\bm{a}}$ around the skyrmion center, taken over the polar angle~$\phi$, for the radial, azimuthal, and out-of-plane components:
\begin{eqnarray}
	\label{eq:B-aver}
	\bar{\bm{B}}^{\bm{a}} = \langle B^{\bm{a}}_r \rangle_\phi \bm{e}_r + \langle B^{\bm{a}}_z \rangle_\phi \bm{e}_z, \quad
	\langle \ldots \rangle_\phi \equiv \int\limits_{-\pi}^{\pi} \dfrac{d\phi}{2\pi} \ldots .
\end{eqnarray}
Consequently, the angle $\bar{\theta}$ satisfies the same Euler–Lagrange equation~\eqref{eq:ELE_theta_coax} as in the coaxial skyrmion case. The exact solution of this equation can be approximated by the coaxial ansatz given in Eq.~\eqref{eq:coax-ansatz}, see also Ref.~\cite{Apostoloff2023}:
\begin{equation}
	\bar{\theta}(r) \approx \theta_{R\delta}^{\gamma \bm{a}}(r) \equiv \theta_{R\delta}(r) + \gamma \theta_{\bar{\bm{b}}^{\bm{a}}}(r) \cos \theta_{R\delta}(r),
	\label{eq:bar-ansatz}
\end{equation}
where $\theta_{R\delta}(r)$ is the domain wall ansatz from Eq.~\eqref{eq:ansatz_free}, and $\theta_{\bar{\bm{b}}^{\bm{a}}}(r)$ is the solution of Eq.~\eqref{eq:thetagamma-ODE} with $b_r = \bar{b}_r^{\bm{a}}$. Recall that $\bm{b}^{\bm{a}} = \bm{B}^{\bm{a}} / B_0$ and $\bar{\bm{b}}^{\bm{a}} = \bar{\bm{B}}^{\bm{a}} / B_0$, where $B_0$ is the characteristic amplitude of the external magnetic field, defined for the Pearl vortex in Eq.~\eqref{eq:gamma-def}.

At the next step of constructing the ansatz, one needs to incorporate the deformation of the skyrmion relative to the radially symmetric magnetization~$\bar{\bm{m}}$. As previously mentioned, this deformation is expected to be small and proportional to the small parameter~$\gamma$ associated with the external field. Therefore, the magnetization of the skyrmion takes the form:
\begin{equation}
	\label{eq:ansatz_eccentric1}
	\bm{m} \approx \bar{\bm{m}} + \gamma \tilde{\bm{m}}.
\end{equation}

Within the linear approximation in~$\gamma$, the deformation vector~$\tilde{\bm{m}}$ must be orthogonal to~$\bar{\bm{m}}$, since both~$\bm{m}$ and~$\bar{\bm{m}}$ are unit vectors. Thus, $\tilde{\bm{m}}$ can be written as a cross product, $\tilde{\bm{m}} = \bar{\bm{m}} \times \bm{\omega}$, where $\bm{\omega}$ can be determined from the following arguments.

Since~$\bar{\bm{m}}$ is expressed via $\bar{\theta}$ from Eq.~\eqref{eq:bar-ansatz}, i.e., it corresponds to a coaxial skyrmion in the averaged field~$\bar{\bm{B}}^{\bm{a}}$, one can use the expression~\eqref{eq:m+mu:coax}. Expanding it to first order in~$\gamma$ yields a linear dependence of~$\bar{\bm{m}}$ on the vector~$\bm{\mu}_{\bar{\bm{b}}^{\bm{a}}} = \theta_{\bar{\bm{b}}^{\bm{a}}} \bm{e}_r$:
\begin{equation}
	\label{eq:for-omega1}
	\bar{\bm{m}} \approx \dfrac{\bm{m}_{R\delta} + \gamma \bm{\mu}_{\bar{\bm{b}}^{\bm{a}}}}{|\bm{m}_{R\delta} + \gamma \bm{\mu}_{\bar{\bm{b}}^{\bm{a}}}|} \approx \bm{m}_{R\delta} + \gamma [\bm{m}_{R\delta} \times \bm{\mu}_{\bar{\bm{b}}^{\bm{a}}} \times \bm{m}_{R\delta}].
\end{equation}
It is natural to expect that the full skyrmion magnetization in linear order in~$\gamma$ depends on the full magnetic field~${\bm{B}}^{\bm{a}}$ in the same way:
\begin{equation}
	\label{eq:for-omega3}
	\bm{m} \approx \dfrac{\bm{m}_{R\delta} + \gamma \bm{\mu}_{\bm{b}^{\bm{a}}}}{|\bm{m}_{R\delta} + \gamma \bm{\mu}_{\bm{b}^{\bm{a}}}|} \approx \bm{m}_{R\delta} + \gamma [\bm{m}_{R\delta} \times \bm{\mu}_{\bm{b}^{\bm{a}}} \times \bm{m}_{R\delta}].
\end{equation}

Returning to Eq.~\eqref{eq:ansatz_eccentric1}, we finally obtain the expression for the deformation part of the magnetization:
\begin{equation}
	\tilde{\bm{m}} = \big[\bar{\bm{m}} \times (\bm{\mu}_{\bm{b}^{\bm{a}}} - \bm{\mu}_{\bar{\bm{b}}^{\bm{a}}}) \times \bar{\bm{m}}\big].
	\label{eq:tilde-m}
\end{equation}

Thus, to linear order in~$\gamma$, the ansatz for the skyrmion magnetization is fully specified by the set of expressions~\eqref{eq:ansatz_eccentric1}, \eqref{eq:bar-m}, and~\eqref{eq:tilde-m}. Note that the ansatz remains valid even up to quadratic order in~$\gamma$ if one adds the term $-\gamma^2 \bar{\bm{m}} \tilde{m}^2 / 2$ on the right-hand side of Eq.~\eqref{eq:ansatz_eccentric1}, see Eq.~\eqref{eq:ansatz_eccentric2}.

Using the ansatz presented here, one can determine the optimal skyrmion parameters — radius~$R$, effective domain wall width~$\delta$, and position~$\bm{a}$ — by minimizing the full free energy~\eqref{eq:F_ferro}. Note that the free energy~\eqref{eq:F_ferro} is a double integral over space, which is computationally expensive to evaluate numerically. In the following subsections, we derive a simplified form of the free energy~\eqref{eq:ferro_expand2}, valid to quadratic order in $\gamma \ll 1$. The corresponding contributions~\eqref{eq:ferro_expand1_explicit} and~\eqref{eq:F2-final} are written as integrals over only the radial coordinate, with the angular integration performed analytically, significantly speeding up numerical computations.

\subsubsection{First-order approximation\label{sec:F1}}

In this subsection, we compute and minimize the free energy up to terms linear in the effective strength~$\gamma \ll 1$ of the external magnetic field. In this case, as indicated in Sec.~\ref{sec:ansatz_ecc}, the leading-order approximation for the skyrmion magnetization~$\bm{m}$ is the radially symmetric unit vector function~$\bar{\bm{m}}$, see Eq.~\eqref{eq:bar-m}. The skyrmion deformation described by the vector~$\tilde{\bm{m}}$ appears only in the first order in~$\gamma$, see Eq.~\eqref{eq:ansatz_eccentric1}.

To determine the unknown skyrmion angle~$\bar{\theta}$ at first order in~$\gamma \ll 1$, we substitute the magnetization~$\bm{m}$ in the form~\eqref{eq:ansatz_eccentric1} into the full free energy~\eqref{eq:F_ferro} and expand it to linear order in~$\gamma$,
\begin{equation}
	\label{eq:ferro_expand1}
	\mathcal{F}_{\text{ferro}}[\bm{m}, \bm{B}^{\bm{a}}] \approx
	\mathcal{F}_{\text{ferro}}[\bar{\bm{m}}, \bm{B}^{\bm{a}}] + \gamma \mathcal{F}_{\text{magn}}^{(1)}[\bar{\bm{m}}, \tilde{\bm{m}}],
\end{equation}
where $\mathcal{F}_{\text{magn}}^{(1)}$ is the first variation of the magnetic energy from Eq.~\eqref{eq:MagFe},
\begin{eqnarray}
 	&&\mathcal{F}_{\text{magn}}^{(1)}[\bar{\bm{m}}, \tilde{\bm{m}}]
 	= -2 d_F \int d^2 \bm{r} \left[ A \tilde{\bm{m}} \cdot \Delta \bar{\bm{m}} + K \tilde{m}_z \bar{m}_z 
 	\right.
 	\notag\\
 	&&\qquad\qquad \left. {} - D \left( \tilde{m}_z \nabla \cdot \bar{\bm{m}} - \tilde{\bm{m}} \cdot \nabla \bar{m}_z \right) \right].
\end{eqnarray}

Note that integration over the polar angle~$\phi$ in the above expression is only required for the radial and out-of-plane components of~$\tilde{\bm{m}}$, since $\bar{\bm{m}}$ is radially symmetric. This implies that if the angular average of these components is zero,
\begin{equation}
	\label{eq:aver_tilde_m}
	\langle \tilde{m}_r \rangle_\phi = \langle \tilde{m}_z \rangle_\phi = 0,
\end{equation}
then the first variation $\mathcal{F}_{\text{magn}}^{(1)}$ will vanish as well. Additionally, since $\tilde{\bm{m}}$ must be orthogonal to~$\bar{\bm{m}}$ and can be written as $\tilde{\bm{m}} = \bar{\bm{m}} \times \bm{\omega}$, this condition simplifies to $\langle \omega_\phi \rangle_\phi = 0$.

On the other hand, if we shift $\bar{\theta} \to \bar{\theta} + \gamma \vartheta$ in Eq.~\eqref{eq:ansatz_eccentric1}, then $\bm{\omega}$ should be changed as $\omega_\phi \to \omega_\phi + \vartheta$ to keep $\bm{m}$ unchanged. To resolve this ambiguity, we may fix the function $\omega_\phi$ such that its angular average vanishes, i.e., $\langle \omega_\phi \rangle_\phi = 0$, and define $\bar{\theta}$ only up to terms linear in $\gamma \ll 1$. This condition for ${\bm{\omega} = [ (\bm{\mu}_{\bm{b}^{\bm{a}}} - \bm{\mu}_{\bar{\bm{b}}^{\bm{a}}}) \times \bar{\bm{m}} ]}$ from Eq.~\eqref{eq:tilde-m} simplifies to $\langle \mu_r^{\bm{a}} \rangle_\phi = 0$, where
\begin{equation}
	\bm{\mu}^{\bm{a}} = \bm{\mu}_{\bm{b}^{\bm{a}}} - \bm{\mu}_{\bar{\bm{b}}^{\bm{a}}},
	\label{eq:omega2}
\end{equation}
and %this follows from
is obvious due to
\begin{equation}
	\langle \bm{\mu}_{\bm{b}^{\bm{a}}} \cdot \bm{e}_r \rangle_\phi = \bm{\mu}_{\bar{\bm{b}}^{\bm{a}}} \cdot \bm{e}_r = \theta_{\bar{\bm{b}}^{\bm{a}}}(r),
\end{equation}
%as
which 
can be verified by comparing Eq.~\eqref{eq:thetagamma-ODE} for ${\bm{b} = \bar{\bm{b}}^{\bm{a}}}$ with Eq.~\eqref{eq:mu_b_ELE} for $\bm{b} = \bm{b}^{\bm{a}}$ after multiplication by $\bm{e}_r$ and angular averaging.

As a result of this choice, only the first term $\mathcal{F}_{\text{ferro}}[\bar{\bm{m}}, \bm{B}^{\bm{a}}]$ in Eq.~\eqref{eq:ferro_expand1} is retained. Since $\bar{\bm{m}}$ is radially symmetric, angular integration allows replacing the external magnetic field~$\bm{B}^{\bm{a}}$ with its angular average~$\bar{\bm{B}}^{\bm{a}}$, see Eq.~\eqref{eq:B-aver}. Consequently, the full free energy to linear order in~$\gamma \ll 1$ becomes:
\begin{eqnarray}
	&&
	\dfrac{\mathcal{F}_{\text{ferro}}[\bm{m}, \bm{B}^{\bm{a}}]}{2\pi d_F A}
	\approx
	\dfrac{\mathcal{F}_{\text{ferro}}[\bar{\bm{m}}, \bar{\bm{B}}^{\bm{a}}]}{2\pi d_F A}
	=
	\dfrac{\mathcal{F}_{\text{ferro}}[\bm{m}_{\bm{b}}, \bm{B}]}{2\pi d_F A}
	+{}
	\notag
	\\ \notag
	&&
	{}+ \int\limits_0^\infty \dfrac{dr \, r}{\ell_{w}^{2}} \left\{ 
	\left( \frac{\ell_{w}^2}{r^2} + 1 \right)\sin^2\bar{\theta}
	+ 2\epsilon \left( \ell_{w} \partial_r \bar{\theta} + \frac{\sin 2\bar{\theta}}{2r/\ell_{w}} \right)
	\right.
	\\
	&& \left. \qquad
	{} + \ell_{w}^{2} (\partial_r \bar{\theta})^2
	- 2\gamma \left[ \bar{b}_r^{\bm{a}} \sin \bar{\theta}
	- 2 \bar{b}_z^{\bm{a}} \sin^2(\bar{\theta}/2) \right]
	\right\}.
	\label{eq:ferro_expand1_explicit}
\end{eqnarray}
Here, $\mathcal{F}_{\text{ferro}}[\bm{m}_{\bm{b}}, \bm{B}]$ is the free energy of the ``no-skyrmion'' background configuration, which does not depend on any skyrmion parameters such as the radius~$R$, domain wall width~$\delta$, or position~$\bm{a}$, and can thus be omitted during minimization.

Minimizing the free energy~\eqref{eq:ferro_expand1_explicit} with respect to the skyrmion angle~$\bar{\theta}(r)$ yields the Euler–Lagrange equation,
\begin{eqnarray}
	\label{eq:ELE_theta_bar}
	&&\dfrac{\ell_{w}^2}{r} \partial_r (r \partial_r \bar{\theta}) - \frac{\ell_{w}^2 + r^2}{2 r^2} \sin 2\bar{\theta}
	+ 2\epsilon \frac{\sin^2 \bar{\theta}}{r/\ell_{w}} =
	\notag \\
	&& \qquad\qquad = \gamma \left( \bar{b}_z^{\bm{a}} \sin \bar{\theta} - \bar{b}_r^{\bm{a}} \cos \bar{\theta} \right),
\end{eqnarray}
which is identical in form to Eq.~\eqref{eq:ELE_theta_coax} for coaxial skyrmions, but with~$\bar{b}_{r/z}^{\bm{a}}$ replacing~$b_{r/z}$. This equation must be supplemented with boundary conditions analogous to Eq.~\eqref{eq:bound_theta0}, with corresponding physical meaning:
\begin{equation}
	\bar{\theta}(r \to \infty) = 0, \quad
	\bar{\theta}(r = 0) = \chi\pi.
	\label{eq:bound_theta_bar}
\end{equation}

Note that the exact solution of Eq.~\eqref{eq:ELE_theta_bar} can be accurately approximated using the ansatz~\eqref{eq:bar-ansatz}, just as in the coaxial case described in Sec.~\ref{sec:ansatz_coax}, by substituting $\bm{b} \to \bar{\bm{b}}^{\bm{a}}$.

Finally, we emphasize that the free energy expression~\eqref{eq:ferro_expand1_explicit} can be formally minimized not only with respect to~$\bar{\theta}$, but also with respect to the skyrmion position~$\bm{a}$. However, it can be shown that the resulting~$\bm{a}$ is accurate only to zeroth order in the small parameter~$\gamma$, see Ref.~\cite{Apostoloff2024} for details. This is because the skyrmion position is highly sensitive not only to its size, determined by the parameters~$R$ and~$\delta$, but also to the shape deformation caused by the external field, which appears in both~$\bar{\theta}$ and~$\tilde{\bm{m}}$. To properly account for this and determine~$\bm{a}$ more accurately, one must compute the free energy to second order in~$\gamma \ll 1$, which is done in the next subsection.

\subsubsection{Second-order approximation\label{sec:F2}}

To accurately determine the dependence of the skyrmion position~$\bm{a}$ on the effective field strength~$\gamma \ll 1$, it is necessary to evaluate the total energy $\mathcal{F}_{\text{ferro}}[\bm{m}, \bm{B}^{\bm{a}}]$ up to second order in~$\gamma$, and then minimize it with respect to~$\bm{a}$. For this purpose, we must extend the expansion of~$\bm{m}$ to include second-order terms in~$\gamma$. Formally, we add a term $\gamma^2 \bm{\lambda}$ to Eq.~\eqref{eq:ansatz_eccentric1}, where $\bm{\lambda}$ is an arbitrary vector function of order~$\gamma^0$, and then normalize $\bm{m}$ to unit length. This yields the following expansion up to order~$\gamma^2$:
\begin{equation}
	\label{eq:ansatz_eccentric2}
	\bm{m} \approx \bar{\bm{m}} + \gamma \tilde{\bm{m}}
	- \frac{\gamma^2}{2} \bar{\bm{m}} \tilde{m}^2
	+ \gamma^2 \left[ \bar{\bm{m}} \times \bm{\lambda} \times \bar{\bm{m}} \right].
\end{equation}

Substituting this form of $\bm{m}$ into Eq.~\eqref{eq:F_ferro}, we obtain the total energy:
\begin{equation}
	\label{eq:ferro_expand2}
	\mathcal{F}_{\text{ferro}}[\bm{m}, \bm{B}^{\bm{a}}] \approx
	\mathcal{F}_{\text{ferro}}[\bar{\bm{m}}, \bar{\bm{B}}^{\bm{a}}] + \mathcal{F}^{(2)}[\bar{\bm{m}}, \tilde{\bm{m}}, \bm{a}],
\end{equation}
where the first term is defined in Eq.~\eqref{eq:ferro_expand1_explicit}, and the second-order correction takes the form:
\begin{eqnarray}
	\label{eq:F2}
	&&\mathcal{F}^{(2)}[\bar{\bm{m}},\tilde{\bm{m}},\bm{a}]={}
	\notag
	\\&&\qquad{}
	=\gamma^2\mathcal{F}_{\text{ferro}}'[\tilde{\bm{m}},\bm{B}^{\bm{a}}]
	-\gamma^2\mathcal{F}_{\text{magn}}^{(1)}[\bar{\bm{m}},\bar{\bm{m}} \tilde{m}^2/2].
	\qquad
\end{eqnarray}
Here, the prime in the first term indicates that when computing $\mathcal{F}_{\text{magn}}$ according to Eq.~\eqref{eq:MagFe}, the constant term~$K$ in the integrand of the full free energy~\eqref{eq:F_ferro} is excluded.

Among the $\gamma^3$ and higher-order contributions omitted in Eq.~\eqref{eq:F2}, one particular term deserves attention:
\begin{gather}
	\dfrac{\gamma^2 \mathcal{F}_{\text{magn}}^{(1)} \left\{ \bar{\bm{m}}, \left[ \bar{\bm{m}} \times \bm{\lambda} \times \bar{\bm{m}} \right] \right\} }{2\pi d_F A} = 
	2\gamma^2 \int \frac{dr\,r}{\ell_{w}^2} [\bm{\lambda} \times \bar{\bm{m}}]_\phi 
	\notag\\
	\times \left[
	\dfrac{\ell_{w}^2}{r} \partial_r(r \partial_r \bar{\theta}) - \frac{\ell_{w}^2 + r^2}{2r^2} \sin 2\bar{\theta}
	+ 2\epsilon \frac{\sin^2 \bar{\theta}}{r/\ell_{w}}
	\right].
	\label{eq:F2-spare1}
\end{gather}
At first glance, this term appears to be of order~$\gamma^2$, but the expression in square brackets is actually of order~$\gamma$ because it corresponds to the left-hand side of Eq.~\eqref{eq:ELE_theta_bar}. Therefore, the entire term is of order~$\gamma^3$ and can be neglected. Consequently, 
%determining 
the determination of
the unknown vector function~$\bm{\lambda}$ is not required to compute the free energy up to order~$\gamma^2$.

It is also important to note that~$\bar{\theta}$, which enters the expression~$\mathcal{F}_{\text{ferro}}[\bar{\bm{m}}, \bar{\bm{B}}^{\bm{a}}]$ in Eq.~\eqref{eq:ferro_expand2}, should be calculated only to first order, as given by Eqs.~\eqref{eq:ELE_theta_bar} and~\eqref{eq:bar-ansatz}. Any corrections to $\bar{\theta}$ of order~$\gamma^2$ can be interpreted as modifications 
%to
of~$\bm{\lambda}$ in Eq.~\eqref{eq:ansatz_eccentric2}, and thus contribute to the energy at order~$\gamma^3$, making them negligible.

Now, using Eq.~\eqref{eq:tilde-m} for $\tilde{\bm{m}}$, we can perform angular averaging over~$\phi$ in Eq.~\eqref{eq:F2} and derive an explicit expression for $\mathcal{F}^{(2)}$:
\begin{gather}
	\dfrac{\mathcal{F}^{(2)}[\bar{\bm{m}}, \tilde{\bm{m}}, \bm{a}]}{2\pi d_F A} =
	\gamma^2 \int \frac{dr\,r}{\ell_{w}^2} \left\{ 
	2 \tilde{b}_r^{\bm{a}} \sin^2 \bar{\theta}
	+ \tilde{b}_z^{\bm{a}} \sin 2\bar{\theta}
	\right.
	\notag\\
	+ \left[
	\tilde{\mu}_r^{\bm{a}} \sin^2 \bar{\theta} - \tilde{\mu}_\phi^{\bm{a}} 
	\right] \ell_w^2 (\partial_r \bar{\theta})^2
	- \frac{\tilde{\mu}_r^{\bm{a}}}{2r^2} (\ell_w^2 + r^2) \sin^2 2\bar{\theta}
	\notag\\
	\left.
	- 2\epsilon \left[
	\tilde{\mu}_\phi^{\bm{a}} \ell_w \partial_r \bar{\theta}
	+ \frac{\tilde{\mu}_{r\times\phi}^{\bm{a}}}{2r/\ell_w} \sin 2\bar{\theta}
	+ \frac{\tilde{\mu}_r^{\bm{a}}}{4r/\ell_w} \sin 4\bar{\theta}
	\right]
	\right\}.
	\label{eq:F2-final}
\end{gather}
Here the following auxiliary functions are introduced, which result from the external magnetic field and depend on the skyrmion position~$\bm{a}$:
\begin{gather}
	\tilde{b}_r^{\bm{a}}(r)=
	\big\langle{\mu}^{\bm{a}}_r\cdot(b_{r}^{\bm{a}}+ 
	\ell_{w}^2\Delta{\mu}^{\bm{a}}_r/2)\big \rangle_\phi
	-(\ell_{w}/r)^2\tilde{\mu}^{\bm{a}}_{r\times\phi}+{}
	\qquad
	\notag\\
	\qquad\qquad\qquad\qquad
	{}
	+[(\ell_{w}/r)^2-1][\tilde{\mu}^{\bm{a}}_{r}   +\tilde{\mu}^{\bm{a}}_{\phi}   ]/2
	,\label{eq:tilde-func}
	\\\notag
	\tilde{b}_z^{\bm{a}}(r) = \left\langle \mu_r^{\bm{a}} b_z^{\bm{a}} \right\rangle_\phi, 
	\;
	\tilde{\mu}_r^{\bm{a}}(r) = \left\langle (\mu_r^{\bm{a}})^2 \right\rangle_\phi, 
	\;
	\tilde{\mu}_\phi^{\bm{a}}(r) = \left\langle (\mu_\phi^{\bm{a}})^2 \right\rangle_\phi,
	\\\notag
	\tilde{\mu}_{r\times\phi}^{\bm{a}}(r) = \tilde{\mu}_r^{\bm{a}} + \tilde{\mu}_\phi^{\bm{a}} +
	\left\langle \mu_r^{\bm{a}} \partial_\phi \mu_\phi^{\bm{a}} - \mu_\phi^{\bm{a}} \partial_\phi \mu_r^{\bm{a}} \right\rangle_\phi.
\end{gather}

\subsection{Optimal positions of the skyrmion and the vortex\label{sec:results_ecc}}

In this subsection, we present the results of applying the previously described variational approach to study stable eccentric configurations of a N\'eel-type skyrmion subjected to the stray field of a Pearl vortex. A comparison with micromagnetic simulations is also provided.

It is important to note that the obtained results indicate that eccentric configurations can only be realized for skyrmions with positive chirality, $\chi = +1$. For a nearly free skyrmion ($\gamma \to 0$), this phenomenon was predicted in Ref.~\cite{Andriyakhina2021}. Let us provide a qualitative explanation. In the limit $\gamma \to 0$, the skyrmion's shape is independent of the distance~$a$ to the vortex center, and its free energy for chirality~$\chi$ can be written as $\mathcal{F} = \mathcal{F}_+ - \chi \mathcal{F}_-$, where $\mathcal{F}_\pm$ are monotonic functions of the distance. When $\chi = -1$, the function $\mathcal{F}(a)$ is monotonic, and the only minimum is at $a = 0$. In contrast, for $\chi = +1$, the subtraction of two monotonic functions may yield a function with multiple minima. This qualitative behavior persists for arbitrary~$\gamma$, although the skyrmion’s size and shape begin to depend on the distance.

To minimize the total free energy~\eqref{eq:ferro_expand2} with the corresponding contributions~\eqref{eq:ferro_expand1_explicit} and~\eqref{eq:F2-final}, one must first evaluate the functions associated with the shifted vortex field~$\bm{B}^{\bm{a}}(\bm{r}) = \bm{B}_V(\bm{r}_{\bm{a}})$, where ${\bm{r}_{\bm{a}} = \bm{r} + \bm{a}}$, see Eq.~\eqref{eq:vortex:B} and the explanation about the origin of the shift in Subsection~\ref{sec:ansatz_ecc}. Under the assumptions~\eqref{eq:assumptions}, some of these functions can be evaluated analytically:
\begin{eqnarray}
\label{eq:thetagamma-approx}
\bm{b}^{\bm{a}}&=& -\dfrac{\bm{e}_{\bm{r}_{\bm{a}}}+\bm{e}_z}{r_{\bm{a}}/\ell_{w}},   
\quad
\bm{\mu}_{\bm{b}^{\bm{a}}}=\theta_{\bm{b}}(r_{\bm{a}})\bm{e}_{\bm{r}_{\bm{a}}},
\\
\bar{\bm{b}}^{\bm{a}}&=& \bar{b}_{r}^{\bm{a}}(r)\bm{e}_{r}+\bar{b}_{z}^{\bm{a}}(r)\bm{e}_z,  
\quad
\bm{\mu}_{\bar{\bm{b}}^{\bm{a}}}=\theta_{\bar{\bm{b}}^{\bm{a}}}(r)\bm{e}_{r},
\\
\bar{b}_{r}^{\bm{a}}(r)&=& -\frac{\Theta(r-a)}{r/\ell_{w}},
\quad
\bar{b}_{z}^{\bm{a}}(r)=-\frac{K[4 a r/(a+r)^2] }{\pi  (a+r)/2\ell_{w}},
\qquad
\label{eq:vortex:field0}
\\
\label{eq:thetaAgamma-approx}
\theta_{\bar{\bm{b}}^{\bm{a}}}(r)&=&\big[I_0(a/\ell_{w})K_1(r/\ell_{w})-\ell_{w}/r\big]\Theta(r-a)
\notag\\&&{}
-K_0(a/\ell_{w})I_1(r/\ell_{w})\Theta(a-r), 
\end{eqnarray}
also see Eqs.~\eqref{eq:vortex:h:approx} and~\eqref{eq:theta_b:vortex}.
Here, $\bm{e}_{\bm{r}_{\bm{a}}} = \bm{r}_{\bm{a}} / r_{\bm{a}}$ is the unit radial vector with respect to the shifted vortex center. The function $I_n(z)$ denotes the modified Bessel functions of the first kind, $K[z]$ is the complete elliptic integral of the first kind, and $\Theta(z)$ is the Heaviside step function.

The remaining functions, $\tilde{b}_{r/z}^{\bm{a}}$, $\tilde{\mu}_{r/\phi}^{\bm{a}}$, and $\tilde{\mu}_{r\times\phi}^{\bm{a}}$, are computed numerically according to Eq.~\eqref{eq:tilde-func}. Once these are known, the total free energy can be minimized using the full form given in Eq.~\eqref{eq:ferro_expand2}, along with explicit expressions from Eqs.~\eqref{eq:ferro_expand1_explicit} and~\eqref{eq:F2-final}, substituting $\bar{\theta}$ as per Eq.~\eqref{eq:bar-ansatz}.

Due to the radial symmetry of the Pearl vortex, the skyrmion position~$\bm{a}$ enters the total free energy only through its magnitude~$a$, i.e., the distance between the centers of the skyrmion and the vortex. Thus, the minimization is carried out over three skyrmion parameters: the radius~$R$, the effective domain wall width~$\delta$, and the distance~$a$ from the vortex center. The stable solution with~$a=0$ corresponds to the coaxial configuration discussed in Section~\ref{Sec:Coax} and Ref.~\cite{Apostoloff2023}. Eccentric configurations of the skyrmion, where $a \neq 0$, are discussed below.

\subsubsection{Skyrmion radius and distance in the eccentric configuration\label{sec:aRvsgamma}}

\begin{figure}[t]
	\includegraphics[width=0.45\textwidth]{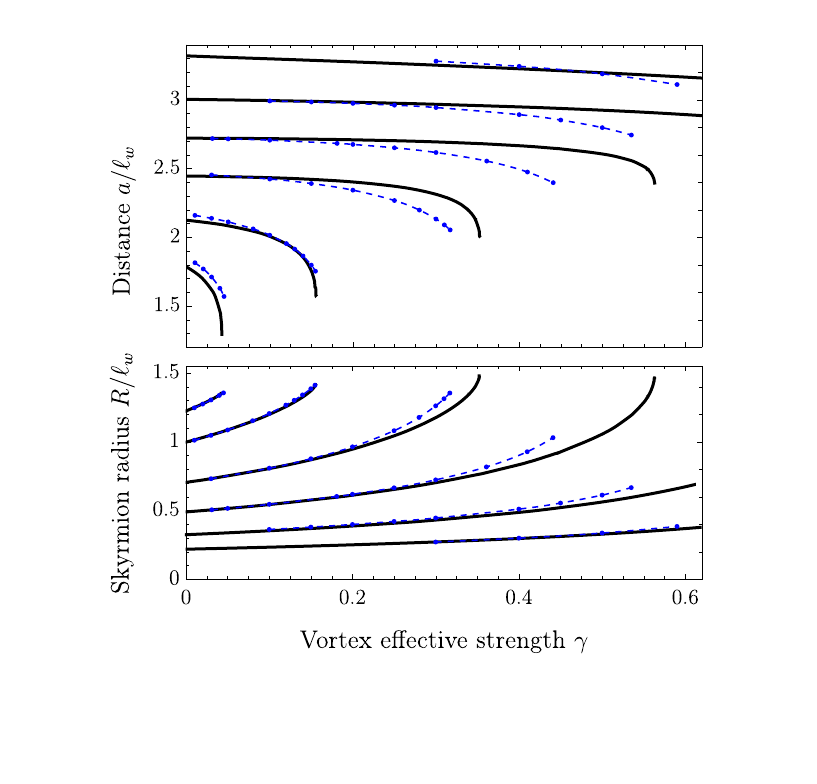}
	\caption{Dependence of the skyrmion radius $R/\ell_w$ (lower panel) and the distance $a/\ell_w$ between the skyrmion and vortex centers (upper panel) on the effective vortex strength $\gamma$, obtained by minimizing the total free energy, Eq.~\eqref{eq:ferro_expand2}, using the ansatz for $\bar{\theta}$, Eq.~\eqref{eq:bar-ansatz}. The different curves correspond to various DMI parameters: $\epsilon = 0.25$, $0.3$, $0.35$, $0.4$, $0.45$, and $0.48$ (top to bottom in the upper panel and bottom to top in the lower panel). Circles connected by dashed lines indicate analogous dependencies obtained from micromagnetic simulations.}
	\label{fig:aRvsgamma}
\end{figure}

Figure~\ref{fig:aRvsgamma} shows the skyrmion radius~$R$ (lower panel) and the distance~$a$ between the centers of the skyrmion and the vortex (upper panel) in stable eccentric configurations as functions of the effective vortex strength~$\gamma$, for different values of the DMI parameter~$\epsilon$. Solid lines correspond to the results obtained from minimizing Eq.~\eqref{eq:ferro_expand2} using the ansatz~\eqref{eq:bar-ansatz}, while circles represent results from micromagnetic simulations (see Subsection~\ref{sec:MMM}).

%%%%%%%%%%%%%%%%%%%%%%%%%%%%%%
% Figure 6
%%%%%%%%%%%%%%%%%%%%%%%%%%%%%%
%%%%%%%%%%%%%%%%%%%%%%%%%%%%%%%%%%%

All curves start with a gentle slope at small~$\gamma$ and end with a sharp increase in radius~$R$ %!%and 
as well as 
%!%
a rapid decrease in distance~$a$. This behavior corresponds to a square-root-type dependence near a certain critical value~$\gamma_{\rm cr}^+$:
\begin{gather}
	\label{eq:aR_sqrt}
	R(\gamma_{\rm cr}^+) - R(\gamma) \propto a(\gamma) - a(\gamma_{\rm cr}^+) \propto \sqrt{\gamma_{\rm cr}^+ - \gamma},
\end{gather}
for $|\gamma_{\rm cr}^+ - \gamma| \ll \gamma_{\rm cr}^+$.

As discussed in Subsection~\ref{sec:var_appr}, accurate determination of the distance~$a$ critically depends on the second-order contributions in~$\gamma$, see Eq.~\eqref{eq:F2-final}. The precision of these contributions is reflected in the good agreement of the $a(\gamma)$ curves %!%
(obtained with the help of analytical ansatz and micromagnetic modeling)
%!%
in Fig.~\ref{fig:aRvsgamma}, although some quantitative deviation is observed near~$\gamma_{\rm cr}^+$. This arises because the parameters $R$, $\delta$, and $a$ are formally defined only to linear order in~$\gamma$. Nevertheless, Fig.~\ref{fig:aRvsgamma} shows that even near~$\gamma_{\rm cr}^+$ the qualitative (and even quantitative, for $\gamma \ll 1$ and $\epsilon > 0.4$) behavior described by Eq.~\eqref{eq:aR_sqrt} is well reproduced. Moreover, the distance~$a$ is particularly sensitive to small changes in the free energy, so unaccounted effects contributing at order~$\gamma^2$ may affect the experimentally observable separation between the skyrmion and the vortex.

Note also an interesting feature in Fig.~\ref{fig:aRvsgamma}: the skyrmion radius and the distance between the centers of the skyrmion and the vortex converge as~$\gamma$ approaches~$\gamma_{\rm cr}^+$. This feature can be explained physically as follows. As the vortex strength~$\gamma$ increases, the skyrmion radius~$R$ grows due to the influence of the stray magnetic field. A larger radius requires a smaller distance~$a$, which can be confirmed by evaluating the Zeeman energy, Eq.~\eqref{eq:F-Sk-V-0}, as a function of radius for a free skyrmion. When~$R \ll a$, the skyrmion only contacts the vortex center with its “tail,” where the magnetization is approximately vertical, $\bm{m} = \bm{e}_z$. However, when the skyrmion radius~$R$ becomes comparable to the distance~$a$, the domain wall — where the magnetization significantly deviates from the vertical — begins to overlap with the vortex center. This configuration is energetically unfavorable, and the skyrmion relocates directly above the vortex core, meaning the eccentric configuration transitions into a coaxial one.

\subsubsection{Free energy vs.\ distance: multiple minima\label{sec:Fa}}

Let us analyze the dependence of the free energy on the distance~$a$ in detail to determine which configuration—eccentric or coaxial—is energetically favorable for given values of~$\epsilon$ and~$\gamma$. To this end, we define the function~$\mathfrak{F}(a)$ as
\begin{equation}
\label{eq:Fa-aver-Rdelta}
\mathfrak{F}(a)\equiv \min_{R,\delta} \dfrac{\mathcal{F}_{\text{ferro}}[\bm{m},\bm{B}^{\bm{a}}]}{2\pi d_F A},
\end{equation}
which represents the total free energy $\mathcal{F}_{\text{ferro}}[\bm{m},\bm{B}^{\bm{a}}]$ from Eq.~\eqref{eq:ferro_expand2}, with $\bar{\theta}$ taken from Eq.~\eqref{eq:bar-ansatz}, normalized by $2\pi d_F A$ and minimized with respect to the skyrmion's geometric parameters: radius~$R$ and effective domain wall width~$\delta$. A minimum $\mathfrak{F}(a)$ at $a=a_{\rm min}$ indicates a possible stable skyrmion configuration in the stray field of a Pearl vortex.

\begin{figure}[t]
\includegraphics[width=0.45\textwidth]{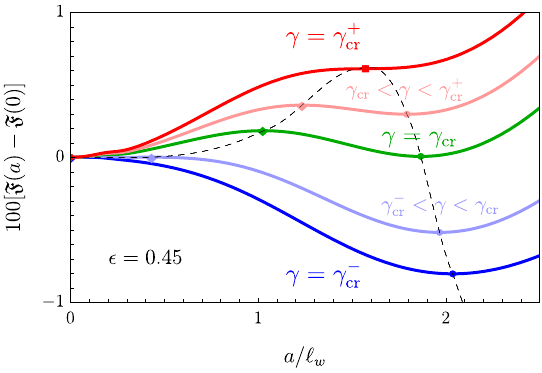}
\caption{Plot of $100[\mathfrak{F}(a)-\mathfrak{F}(0)]$, obtained from the total free energy, Eq.~\eqref{eq:Fa-aver-Rdelta}, for DMI parameter $\epsilon=0.45$ and several values of the effective vortex strength~$\gamma$: $\gamma_{\rm cr}^{-}\approx0.106$, $0.119$, $\gamma_{\rm cr}\approx0.138$, $0.147$, $\gamma_{\rm cr}^{+}\approx0.156$ (bottom to top). Circles, diamonds, and squares indicate minima, maxima, and inflection points, respectively. The dashed curve traces the locations of minima and maxima as $\gamma$ is varied continuously. All plotted functions increase monotonically for $a > 2.5\ell_w$ beyond the visible range.}
\label{fig:F(a)}
\end{figure}

Figure~\ref{fig:F(a)} illustrates the behavior of~$\mathfrak{F}(a)$ for $\epsilon = 0.45$ and several values of~$\gamma$, representing the typical types of behavior classified by three critical values: $\gamma_{\rm cr}^{-}(\epsilon)$, $\gamma_{\rm cr}(\epsilon)$, and $\gamma_{\rm cr}^{+}(\epsilon)$, as discussed below.

%%%%%%%%%%%%%%%%%%%%%%%%%%%%%%
% Figure 7
%%%%%%%%%%%%%%%%%%%%%%%%%%%%%%
%%%%%%%%%%%%%%%%%%%%%%%%%%%%%%%%%%%

For $\epsilon < \epsilon_{\rm cr}^{-} \approx 0.488$ and $\gamma < \gamma_{\rm cr}^{-}(\epsilon)$, the function~$\mathfrak{F}(a)$ has a single minimum at $a = a_{\rm min} > 0$, indicating that the only stable skyrmion--vortex configuration is eccentric.

When $\gamma > \gamma_{\rm cr}^{+}(\epsilon)$ or $\epsilon > \epsilon_{\rm cr}^{+} \approx 0.493$, the function~$\mathfrak{F}(a)$ has a single minimum at $a = 0$, meaning that the coaxial skyrmion--vortex configuration is the only possible stable state.

In the intermediate regime, $\gamma_{\rm cr}^{-}(\epsilon) < \gamma < \gamma_{\rm cr}^{+}(\epsilon)$, the system can support both coaxial and eccentric skyrmion--vortex configurations, as $\mathfrak{F}(a)$ possesses at least two minima: one at $a = 0$ and another at $a = a_{\rm min} > 0$. However, the free energy values for these configurations differ: the global minimum occurs for the coaxial configuration when $\gamma_{\rm cr}(\epsilon) < \gamma < \gamma_{\rm cr}^{+}(\epsilon)$, and for the eccentric configuration when $\gamma_{\rm cr}^{-}(\epsilon) < \gamma < \gamma_{\rm cr}(\epsilon)$.

Remarkably, there is no fundamental restriction on $\mathfrak{F}(a)$ having more than two minima. In particular, we have found an additional local minimum at $a = a_{\rm add} \sim 0.2\ell_w$. However, for all values of~$\epsilon$ and~$\gamma$ considered, this minimum remains local and its small depth falls outside the precision of our second-order expansion in~$\gamma \ll 1$. Consequently, the variational analytical approach alone cannot guarantee the existence of such a stable configuration. Furthermore, micromagnetic simulations did not confirm its presence.

The dashed line in Fig.~\ref{fig:F(a)} shows the evolution of the minima and maxima as $\gamma$ is varied continuously. Near $\gamma \approx \gamma_{\rm cr}^{+}$, the location of the minimum~$a_{\rm min}$ changes rapidly with~$\gamma$, approaching the maximum point~$a_{\rm max}$ and merging at an inflection point $a_{\rm in}$ precisely at $\gamma = \gamma_{\rm cr}^{+}$. This rapid change leads to a sharp decrease (increase) in the distance~$a$ (skyrmion radius~$R$), as shown in Fig.~\ref{fig:aRvsgamma}.

\subsubsection{Phase diagram\label{sec:phase_diag}}

\begin{figure}[t]
\includegraphics[width=0.45\textwidth]{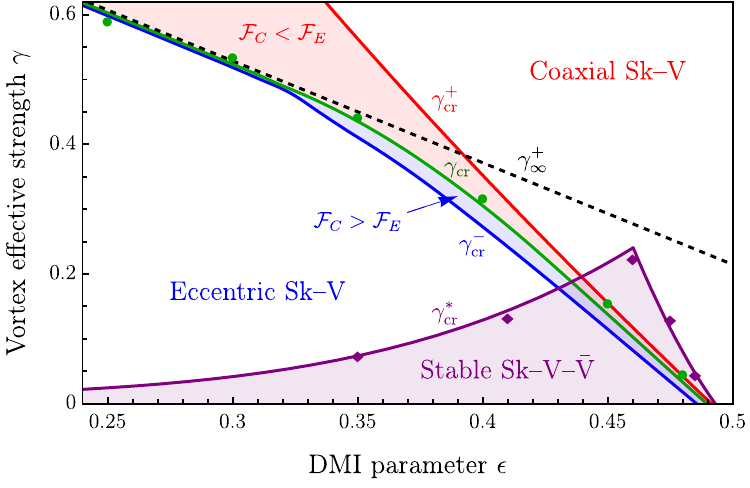}
\caption{Phase diagram based on the total free energy from Eq.~\eqref{eq:ferro_expand2} with the variational ansatz for $\bar{\theta}$ from Eq.~\eqref{eq:bar-ansatz}. The blue, green, and red solid monotonic curves (from bottom to top) represent $\gamma_{\rm cr}^-(\epsilon)$, $\gamma_{\rm cr}(\epsilon)$, and $\gamma_{\rm cr}^+(\epsilon)$, respectively, delineating phase boundaries for skyrmion--vortex configurations. The shaded regions between them denote phases where both coaxial and eccentric states can coexist. The black dashed line denotes $\gamma_{\infty}^+(\epsilon)$, indicating the condition under which the radius of the coaxial skyrmion becomes comparable to the Pearl length, see Eq.~\eqref{eq:gamma_inf}. The violet solid non-monotonic curve shows~$\gamma_{\rm cr}^*(\epsilon)$, which bounds the phase (shaded region below the curve) of stable skyrmion--vortex--antivortex configurations (see Sec.~\ref{sec:VaV-Sk}). Circles and diamonds indicate the values of $\gamma_{\rm cr}$ and $\gamma_{\rm cr}^*$ obtained from micromagnetic simulations.}
\label{fig:phasediag}
\end{figure}

The results of the previous subsection are summarized in the phase diagram on the $(\epsilon, \gamma)$ plane shown in Fig.~\ref{fig:phasediag}. This diagram identifies four distinct phases of skyrmion--vortex configurations (indicated by solid monotonic curves corresponding to the critical values $\gamma_{\rm cr}^-(\epsilon)$, $\gamma_{\rm cr}(\epsilon)$, and $\gamma_{\rm cr}^+(\epsilon)$), as well as the phase of stable skyrmion--vortex--antivortex configurations (shown as a solid non-monotonic line), which will be discussed in more detail in subsection~\ref{sec:VaV-Sk}. The two unshaded regions indicate phases where only eccentric ($\gamma < \gamma_{\rm cr}^-(\epsilon)$) or only coaxial ($\gamma > \gamma_{\rm cr}^+(\epsilon)$) configurations are possible. The solid shaded areas between the curves $\gamma_{\rm cr}^-(\epsilon)$, $\gamma_{\rm cr}(\epsilon)$, and $\gamma_{\rm cr}^+(\epsilon)$ represent phases in which both eccentric and coaxial configurations are theoretically allowed within the same heterostructure, since the free energy exhibits two minima with respect to the distance~$a$. In practice, both configurations were observed in micromagnetic simulations near $\gamma \approx \gamma_{\rm cr}$, as described in subsection~\ref{sec:MMM} and illustrated in Fig.~\ref{fig:MMM}.

%%%%%%%%%%%%%%%%%%%%%%%%%%%%%%
% Figure 8
%%%%%%%%%%%%%%%%%%%%%%%%%%%%%%
%%%%%%%%%%%%%%%%%%%%%%%%%%%%%%%%%%%

Note that the curves $\gamma_{\rm cr}^-(\epsilon)$ and $\gamma_{\rm cr}(\epsilon)$ approach the asymptotic line $\gamma_{\infty}^+(\epsilon) = 1 - \pi\epsilon/2$ for $\epsilon \lesssim 0.3$. As the effective vortex strength $\gamma$ approaches the critical threshold $\gamma_{\infty}^+$, the radius~$R$ of the coaxial skyrmion increases significantly, becoming comparable to the Pearl length $\lambda$, which is much larger than~$\ell_w$; see the details in subsection~\ref{sec:results_coax} and in Ref.~\cite{Apostoloff2023}. Since the free energy of such an enlarged coaxial skyrmion turns out to be substantially lower than that of any eccentric configuration, the value of $\gamma_{\infty}^+$ is always higher than both $\gamma_{\rm cr}^-(\epsilon)$ and $\gamma_{\rm cr}(\epsilon)$, which characterize the transition points between coaxial and eccentric configurations.

Green circles in Fig.~\ref{fig:phasediag} represent the values of $\gamma_{\rm cr}(\epsilon)$ obtained from micromagnetic simulations, confirming the analytical predictions with a high degree of agreement.

\section{Stabilization of a Vortex–Antivortex Pair by a Skyrmion\label{sec:VaV-Sk}}

In this subsection, we discuss a more complex configuration, where a skyrmion is placed in the field of a Pearl vortex–antivortex pair, whose magnetic fields are oriented in opposite directions. We refer to the vortex defined by Eq.~\eqref{eq:vortex:B} as the vortex, and the second one with reversed magnetic field as the antivortex. To define the magnetic field of the antivortex, the sign in Eq.~\eqref{eq:vortex:B} must be reversed:
\begin{eqnarray}
   {\bm B}_{\bar{V}}  = -\phi_0 \sgn(z) \nabla 
   \int \frac{d^2\bm{q}}{(2\pi)^2} \frac{e^{-q |z| +i \bm{q}(\bm{r}-\bar{\bm{a}})}}{q(1+2q\lambda)}.
   \label{eq:anti-vortex:B}
\end{eqnarray}
Here, we have also accounted for the fact that the antivortex is located at $\bm{r} = \bar{\bm{a}}$, while the vortex is assumed to be at the origin, $\bm{r} = 0$; cf. Eq.~\eqref{eq:vortex:B}.

\begin{figure}[t]
 \includegraphics[width=0.45\textwidth]{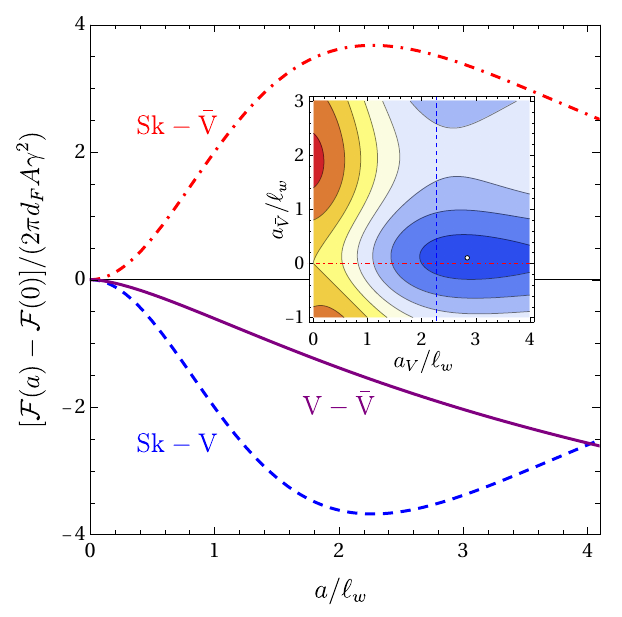}
 \caption{\textit{Main panel:} Normalized free energy $\mathcal{F}$ for vortex–antivortex (solid curve), skyrmion–vortex (dashed curve), and skyrmion–antivortex (dash-dotted curve) pair configurations as a function of distance $a$ between object centers. \textit{Inset:} Total energy (in arbitrary units) of the skyrmion–vortex–antivortex configuration as a function of displacements $a_V$ (vortex) and $a_{\bar{V}}$ (antivortex) relative to the skyrmion. Vertical and horizontal lines indicate energy minima in the pair configurations; the dot marks the actual minimum. Parameters: ${\epsilon=0.41}$, ${\gamma=0.1}$.}
 \label{fig:FVaV_a}
\end{figure}

It is worth noting that the interaction between a skyrmion and a single antivortex is less interesting than the interaction of a skyrmion with a vortex or with a vortex–antivortex pair. The reason is that a skyrmion is generally repelled by an antivortex over a wide range of parameters. Indeed, let us consider very small $\gamma$, when the shape of the skyrmion can be considered unperturbed. As already noted in subsection~\ref{sec:results_ecc}, in this case the skyrmion energy as a function of $a$ is monotonic for negative chirality, $\chi = -1$, and may be non-monotonic for positive chirality, $\chi = +1$. In the first case, the energy decreases monotonically and the skyrmion is repelled by the antivortex for any $a$, meaning that no stationary state can form. In the second case, the skyrmion is attracted to the antivortex only if their centers are sufficiently close, see subsection~\ref{sec:var-Sk-V-aV} and Fig.~\ref{fig:FVaV_a} for details; otherwise, it is also repelled. Moreover, the energy minimum at $a=0$ can be lower than the energy at $a \to \infty$ for $\epsilon < \epsilon_{\rm anti} \approx 0.387$ only. In this work, we do not analyze in detail these coaxial skyrmion–antivortex states and instead proceed to the more interesting skyrmion–vortex–antivortex configurations.

In general, determining the stable configurations of the skyrmion–vortex–antivortex system is similar to the problem solved in the previous subsection. However, there is one essential difference: the additional parameter --- the distance $\bar{a} = |\bar{\bm{a}}|$ between the centers of the vortex and antivortex --- affects not only the energy of the ferromagnet but also that of the superconductor. Therefore, we first describe the simpler problem of determining and analyzing the energy of the ``no-skyrmion'' state, assuming a fixed vortex–antivortex separation $\bar{a}$. Then, based on that result, we will use a variational approach to determine stable skyrmion–vortex–antivortex configurations under variations of all distances between the elements of the configuration.

\subsection{The ``No-Skyrmion'' Configuration in the Magnetic Field of a Vortex--Antivortex Pair}

We begin by considering the case where the distance between the centers of the vortices is comparable to the magnetic length and is small relative to the Pearl length,
\begin{equation}
  	\label{eq:a:VaV}
  	\xi\ll \ell_w\sim\bar{a}\ll\lambda.
\end{equation}
The last condition also explicitly assumes that the magnetic length $\ell_w$ is much greater than the coherence length $\xi$ in the superconductor. Under these conditions, the supercurrents and magnetic field of the vortex–antivortex pair can be treated as a superposition of two individual vortices.

The total free energy of a superconductor–ferromagnet heterostructure containing such a vortex–antivortex pair consists of the sum of the superconducting and ferromagnetic energies:
\begin{equation}
  	\mathcal{F}_{\rm tot}[\bm{m},\bar{\bm{a}}]=\mathcal{F}_{\rm super}[\bar{a}]+\mathcal{F}_{\rm ferro}[\bm{m},{\bm{B}}_V+{\bm B}_{\bar{V}}].
\end{equation}

Assuming that the effective vortex strength $\gamma$ is small, see Eq.~\eqref{eq:gamma-gen-def}, and the magnetization $\bm{m}$ deviates only slightly from the vertical direction, cf. Eq.~\eqref{eq:mu_b},
\begin{equation}
  	\bm{m}=\bm{e}_z+\gamma(\bm{\mu}_{\bm{b}}-\bm{\mu}_{\bm{b}^{\bar{\bm{a}}}}),
\end{equation}
the ferromagnetic free energy can be calculated in leading order in $\gamma$:
\begin{equation}
  	\dfrac{\mathcal{F}_{\rm ferro}[\bm{m},{\bm{B}}_V{+}{\bm B}_{\bar{V}}]}{2\pi d_F A}=
  	-\gamma^2\int \frac{d^2\bm{r}}{2\pi \ell_w^2}(\bm{\mu}_{\bm{b}}{-}\bm{\mu}_{\bm{b}^{\bar{\bm{a}}}})(\bm{b}{-}\bm{b}^{\bar{\bm{a}}}).
  	\label{eq:F_ferro_VaV_0}
\end{equation}
Here $\bm{b}={\bm{B}}_V/B_0$ and $\bm{b}^{\bar{\bm{a}}}=-{\bm B}_{\bar{V}}/B_0$ denote the normalized fields of the vortex and antivortex, respectively, and $\bm{\mu}_{\bm{b}}$, $\bm{\mu}_{\bm{b}^{\bar{\bm{a}}}}$ are the corresponding vectors describing the magnetization tilt due to the magnetic fields; see Eqs.~\eqref{eq:gamma-def}, \eqref{eq:mu_theta}, and~\eqref{eq:thetagamma-approx}. Note that Eq.~\eqref{eq:F_ferro_VaV_0} includes not only the Zeeman energy, but also other 
%!%
necessary
%!%
terms, which reduce the Zeeman contribution exactly by 
half.
%a factor of two.

Under condition~\eqref{eq:a:VaV}, one can use the approximations~\eqref{eq:vortex:h:approx}, \eqref{eq:theta_b:vortex}, and~\eqref{eq:vortex:field0} to obtain an explicit dependence of the energy on $\bar{a}$:
\begin{equation}
  	\dfrac{\mathcal{F}_{\rm ferro}[\bm{m},{\bm{B}}_V{+}{\bm B}_{\bar{V}}]}{2\pi d_F A}=
  	-2  \gamma^2\Big[K_0\Big(\dfrac{\bar{a}}{\ell_w}\Big){+}\ln\dfrac{\bar{a}}{
    {2e^{-\gamma_E}}\ell_w}\Big],
  	\label{eq:F_ferro_VaV_1}
\end{equation}
where $\gamma_E\approx0.577$ is Euler's constant.

The interaction energy of Pearl vortex–antivortex pairs at separation $\bar{a}$ was calculated in Ref.~\cite{Pearl1964}. When the distance is small compared to the Pearl length, this energy equals:
\begin{equation}
  	\mathcal{F}_{\rm super}[\bar{a}]=\dfrac{\phi_0^2}{8\pi^2\lambda}\ln\dfrac{\bar{a}}{\xi},
  	\quad
  	\xi\ll\bar{a}\ll\lambda.
\end{equation}

Normalizing it by the same factor as the magnetic energy, we obtain:
\begin{equation}
  	\dfrac{\mathcal{F}_{\rm super}[\bar{a}]}{2\pi d_F A}=\beta\gamma^2
  	\ln\dfrac{\bar{a}}{\xi},
  	\label{eq:F_super_VaV}
\end{equation}
where the dimensionless parameter $\beta$ is expressed via material parameters of the heterostructure as
\begin{equation}
  	\beta= \dfrac{\phi_0}{2\pi^2 \gamma d_F \ell_{w}M_s }=\dfrac{4 A \lambda}{\pi d_F \ell_{w}^2M_s^2 } ,
\end{equation}
and is small for experimentally relevant values~\cite{Petrovic2021,Machain2021,Xie2024}.

Comparing Eqs.~\eqref{eq:F_ferro_VaV_1} and~\eqref{eq:F_super_VaV}, we conclude that for $\beta \ll 1$ and under condition~\eqref{eq:a:VaV}, the superconducting contribution to the energy can be neglected, and the total energy can be approximated by the ferromagnetic part alone. Note that $\mathcal{F}_{\rm ferro}$, Eq.~\eqref{eq:F_ferro_VaV_1}, is a decreasing function of $\bar{a}$, indicating a repulsion between the vortex and antivortex. Physically, this repulsion arises because the magnetic field of nearby vortices creates an unfavorable magnetization pattern in the ferromagnet.

Importantly, vortices attract when $\bar{a}$ does not satisfy condition~\eqref{eq:a:VaV}. Indeed, for $\bar{a} \ll \ell_w$, the ferromagnetic energy~\eqref{eq:F_ferro_VaV_1} becomes small and can be comparable to or smaller than the superconducting energy~\eqref{eq:F_super_VaV}, causing the vortex and antivortex to annihilate, as would happen in the absence of a ferromagnet.

In the opposite limit $\bar{a} \gg \lambda$, the interaction energy of supercurrents takes the form~\cite{Pearl1964}:
\begin{equation}
  	\dfrac{\mathcal{F}_{\rm super}}{2\pi d_F A}=\mathcal{C}_{\rm super}-2\beta\gamma^2
  	\dfrac{\lambda}{\bar{a}},
  	\label{eq:F_super_VaV_2}
\end{equation}
where $\mathcal{C}_{\rm super}$ is a constant independent of $\bar{a}$. In this regime, the ferromagnetic energy is given by
\begin{equation}
  	\dfrac{\mathcal{F}_{\rm ferro}}{2\pi d_F A}=
  	\mathcal{C}_{\rm ferro}+\dfrac{32
    \gamma^2\lambda^3}{\bar{a}^3},
  	\label{eq:F_ferro_VaV_2}
\end{equation}
where $\mathcal{C}_{\rm ferro}$ is also independent of $\bar{a}$. Minimizing the total energy yields the stationary distance between vortex and antivortex $\bar{a}_{\rm min}=4\lambda/\sqrt{\beta/3}$, which satisfies $\bar{a}_{\rm min}\gg\lambda$.

%%%%%%%%%%%%%%%%%%%%%%%%%%%%%%
% Figure 9
%%%%%%%%%%%%%%%%%%%%%%%%%%%%%%
%%%%%%%%%%%%%%%%%%%%%%%%%%%%%%%%%%%

% Stas: stopped here

\subsection{Variational approach for the skyrmion--vortex--antivortex configuration\label{sec:var-Sk-V-aV}}

Let us now consider a configuration consisting of a vortex--antivortex pair and a skyrmion with positive chirality, and assume that the system parameters are such that this configuration is stable for distances between the centers of the objects comparable to $\ell_w$. As we will see below, there exists a fairly wide range of parameters where this holds, see Fig.~\ref{fig:phasediag}. In Fig.~\ref{fig:FVaV_a}, we show the dependence of the normalized free energy ${[\mathcal{F}(a)-\mathcal{F}(a=0)]/(2\pi d_F A \gamma^2)}$ for three pairwise configurations: vortex--antivortex (solid curve, labeled $\rm V-\bar{V}$), skyrmion--vortex (dashed curve, labeled $\rm Sk-V$), and skyrmion--antivortex (dash-dotted curve, labeled $\rm Sk-\bar{V}$), where $a$ denotes the distance between the centers of the respective objects. From the presented dependencies, it is clear that the vortex and antivortex repel each other, as shown in the previous section. The skyrmion tends to position itself at a certain distance from the center of the vortex, see Subsection~\ref{sec:results_ecc}, and coaxially with the antivortex. Also, for other parameters, the skyrmion may tend to position itself directly above the center of the vortex and repel from the antivortex, but in this case the configuration is obviously not stable.

Since in the considered configuration the skyrmion attracts both the vortex and the antivortex, while the vortex and antivortex repel each other, the centers of all three objects will lie on a single line, with the skyrmion located in the middle between the two vortices. We denote the distances from the skyrmion center to the centers of the vortex and antivortex by~$a_V$ and $a_{\bar{V}}$, respectively, and seek the minimum of the free energy as a function of these two distances, as well as the skyrmion radius~$R$ and effective domain wall width~$\delta$, following the approach developed in Subsection~\ref{sec:var_appr}.

In the inset of Fig.~\ref{fig:FVaV_a}, we show (in arbitrary units) the dependence of the total energy $\mathcal{F}$ of such a configuration as a function of the displacements~$a_V$ of the vortex and $a_{\bar{V}}$ of the antivortex relative to the skyrmion. It can be seen that the minimum of the function, marked by a white dot, is reached at values of~$a_V$ and $a_{\bar{V}}$ that differ from the minima in the corresponding pairwise configurations skyrmion--vortex and skyrmion--antivortex (marked by dashed and dash-dotted lines, respectively). Next, we analyze the stable skyrmion--vortex--antivortex configurations in more detail.

\subsection{Stable configurations}

\begin{figure}[!t]
	\includegraphics[width=0.45\textwidth]{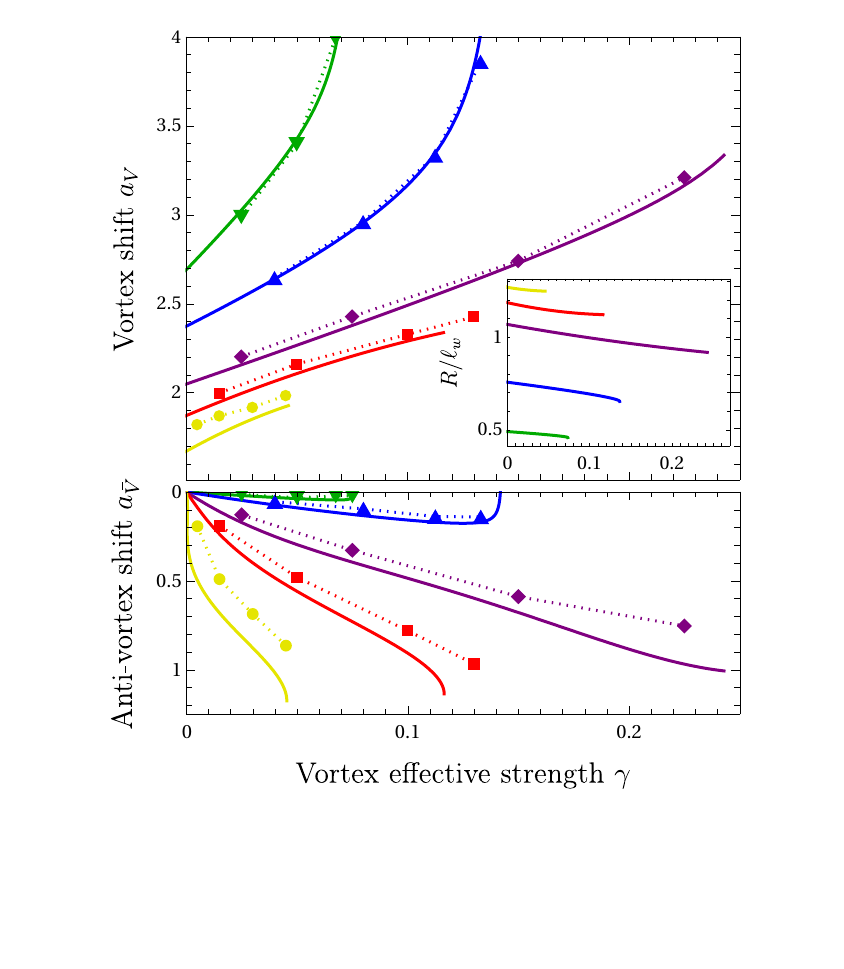}
	\caption{Dependence of the displacements of the vortex center~$a_V$ (top panel) and the antivortex center~$a_{\bar{V}}$ (bottom panel) relative to the skyrmion center, as well as the skyrmion radius~$R$ (inset), on the effective vortex strength~$\gamma$ for several values of the DMI parameter $\epsilon=0.35$; $0.41$; $0.46$; $0.475$; $0.485$ (curves from top to bottom in the main panels and from bottom to top in the inset). Solid curves show the results from the variational approach, and points are extracted from micromagnetic simulations.}
	\label{fig:aVaAV_gamma}
\end{figure}

Figure~\ref{fig:aVaAV_gamma} shows the dependence of the distances~$a_V$ and $a_{\bar{V}}$ from the skyrmion center to the centers of the vortex and antivortex, as well as the skyrmion radius~$R$, as functions of $\gamma$ for several values of the DMI parameter $\epsilon$. The solid curves are obtained using the variational approach, and the dots are the result of micromagnetic simulations. Two distinct regimes can be identified, which can be qualitatively described as follows:
\begin{enumerate}
    \item For $\epsilon<\epsilon_{\rm cr}^*\approx0.46$, the vortex moves away from the skyrmion–antivortex pair as $\gamma$ increases.
    \item For $\epsilon>\epsilon_{\rm cr}^*\approx0.46$, the antivortex moves away from the skyrmion–vortex pair as $\gamma$ increases.
\end{enumerate}

%%%%%%%%%%%%%%%%%%%%%%%%%%%%%%
% Figure 10
%%%%%%%%%%%%%%%%%%%%%%%%%%%%%%
%%%%%%%%%%%%%%%%%%%%%%%%%%%%%%%%%%%

In the indicated regimes, once a certain critical value $\gamma>\gamma_{\rm cr}^*(\epsilon)$ is exceeded, the vortex (or antivortex) moves away from the skyrmion–antivortex (or skyrmion–vortex) pair to a distance $\bar{a}_{\rm min}=4\lambda/\sqrt{\beta/3}\gg\lambda\gg\ell_w$, see the previous subsection, and its influence on the skyrmion effectively vanishes. However, there is a qualitative difference between these regimes at $\gamma\approx\gamma_{\rm cr}^*(\epsilon)$.

In the first regime, the vortex moves away from the skyrmion–antivortex pair gradually, i.e., the distance~$a_V$ can take arbitrary intermediate values depending on $\gamma$.

In the second regime, the antivortex moves away from the skyrmion–vortex pair only within limited bounds, i.e., the distance~$a_{\bar{V}}$ is bounded from above by a value~$a_{\bar{V},\rm cr}$ of the order of $\ell_w$, corresponding to a maximum of the energy as a function of the distance~$a_{\bar{V}}$. The presence of this maximum can be understood as follows. For small $\gamma$, where the change in the skyrmion size can be neglected, the energy dependence on~$a_{\bar{V}}$ enters only through two terms: $\mathcal{F}_{\rm \bar{V}-Sk}(a_{\bar{V}})$ and $\mathcal{F}_{\rm {V}-\bar{V}}(a_{V}+a_{\bar{V}})$. The first of these energies is of order $\gamma$ and reaches a maximum at a finite value of~$a_{\bar{V}}$, while the second is of order $\gamma^2$ and decreases monotonically with increasing~$a_{\bar{V}}$, see Fig.~\ref{fig:FVaV_a}. This means that their sum also has a maximum. As $\gamma$ increases, the role of the second term becomes more significant, and, in addition, the dependence on~$a_{\bar{V}}$ appears via the change in the size and shape of the skyrmion. However, the maximum persists up to $\gamma_{\rm cr}^*$. For larger values of $\gamma$, the maximum in the energy dependence on~$a_{\bar{V}}$ disappears, and the antivortex is forced to jump directly to the distance $\bar{a}_{\rm min}\gg\lambda\gg\ell_w$, skipping intermediate positions.

In the lower part of Fig.~\ref{fig:phasediag}, the region of parameters $(\epsilon,\gamma)$ is shown where stable skyrmion–vortex–antivortex configurations can exist. It is easy to see that the region has a characteristic triangular shape, where the two sloped lines correspond to the two regimes described above. Diamonds mark the critical values $\gamma_{\rm cr}^*$ extracted from micromagnetic simulations. It is seen that the analytical prediction agrees well with the simulation results.

\section{Higher-order skyrmions in the magnetic field of a Pearl vortex\label{sec:high_sk-v}}

In this subsection, following Refs.~\cite{Shustin2024en, Shustin2025en}, it is shown that the stray fields of a superconducting Pearl vortex can give rise to stable configurations involving higher-order skyrmions (HOS) due to the orbital effects of the inhomogeneous magnetic field. The skyrmions considered in the previous sections possessed a topological charge $|Q|=1$, see the definition of the topological charge in Eq.~\eqref{eq:rho_Q}; here, we focus on HOS characterized by $n=|Q|\geqslant2$, see Eq.~\eqref{eq:sk_high}.

\subsection{Three-spin interaction}

To study such states, one must take into account an additional contribution to the free energy of the ferromagnetic film, caused by the so-called scalar chiral or three-spin interaction~\cite{Shustin2023}:
\begin{equation}
  	\mathcal{F}_{K}=\int d^2\bm{r}\,\rho(\bm{r})\mathcal{K}(|\bm{r}-\bm{a}|),\label{eq:FK}
\end{equation}
where $\rho(\bm{r})$ denotes the topological charge density, see Eq.~\eqref{eq:rho_Q}, and the explicit form of the kernel $\mathcal{K}$,
\begin{equation}\label{eq:Hb}
	\mathcal{K}(|\bm{r}-\bm{a}|) = \mathcal{K}_0\sin[\pi\phi_\Delta(|\bm{r}-\bm{a}|)/\phi_0],
\end{equation}
can be derived from the Hubbard model in the strong correlation regime within third-order perturbation theory accounting for orbital effects~\cite{Shustin2023}. Here, $\mathcal{K}_0$ is a constant, $\phi_{\Delta}$ is the magnetic flux of the $B_z$ component of the Pearl vortex field through an elementary plaquette of the triangular ferromagnetic lattice, and the vortex center is located at the point $\bm{a}$. Almost everywhere, except near the vortex core, the argument of the $\sin$ in Eq.~\eqref{eq:Hb} is small and the sine can be replaced by its argument. The influence of the vortex core region turns out to be negligible, as shown by numerical results~\cite{Shustin2024en}, thus one may use the simplified expression implying direct proportionality between the kernel $\mathcal{K}$ and the transverse component $B_z$ of the magnetic field: $\mathcal{K}\propto B_z$.

Substituting the magnetization of a higher-order skyrmion, given by Eq.~\eqref{eq:sk_high} for $n\geqslant2$, into Eqs.~\eqref{eq:MagFe},~\eqref{eq:F-Sk-V-0}, and~\eqref{eq:FK}, representing the energies of the free ferromagnet, Zeeman interaction, and the three-spin interaction respectively, and performing the azimuthal integration around the skyrmion center, one obtains:
\begin{eqnarray}
	\dfrac{\mathcal{F}_{\rm magn}}{2\pi d_F A}&=&\int\limits_0^\infty \dfrac{dr \, r}{\ell_{w}^{2}} \Big[ 
	\Big(\frac{n^2\ell_{w}^2}{r^2}+  1\Big)\sin^2\theta
	+  \ell_{w}^{2}(\partial_r{\theta})^2\Big],
	\quad
	\label{eq:Fmagn2}
	\\
	\label{eq:FZ2}
	\dfrac{\mathcal{F}_{Z}}{2\pi d_F A}&=&
	-2\gamma \int\limits_0^\infty \dfrac{dr \, r}{\ell_{w}^{2}}
	\big[\bar{b}_{r,n}^{\bm{a}}\sin\theta
	-2\bar{b}_z^{\bm{a}}
	\sin^2(\theta/2)
	\big],
	\\
	\label{eq:FK2}
	\dfrac{\mathcal{F}_{K}}{2\pi d_F A}&=&-\kappa\gamma n\int\limits_0^{\infty}dr\,\bar{b}_{z}^{\bm{a}}(r) \partial_r[\cos\theta(r)],
\end{eqnarray}  
Here, $\bar{b}_{z}^{\bm{a}}(r)$ is the $z$-component of the normalized magnetic field of a vortex shifted to point $\bm{a}$ and averaged over the azimuthal angle $\phi$ around the skyrmion center, see Eq.~\eqref{eq:vortex:field0}, while $\bar{b}_{r,n}^{\bm{a}}(r,\phi_0)$ denotes the corresponding radial projection averaged with the weight $\cos\phi^{(n)}$, see Eq.~\eqref{eq:sk_high},
\begin{equation}
	\bar{b}_{r,n}^{\bm{a}}(r,\phi_0)=(-1)^{n}\frac{\cos(n-1)\phi_0}{(r/\ell_w)(r/a)^{n-1}}
	\Theta(r-a).
\end{equation}
The dimensionless parameter $\kappa$ in Eq.~\eqref{eq:FK2} is given by
\begin{equation}
	\kappa=\dfrac{\sqrt{3}\pi \mathcal{K}_0 d^2}{2M_s\phi_0d_F},
\end{equation}
where $d$ is the lattice constant of the ferromagnet.

Note that in the expression~\eqref{eq:Fmagn2} for the free energy of the ferromagnet, terms due to the Dzyaloshinskii–Moriya interaction (DMI) are absent. This is because for a radially symmetric skyrmion with $n\geqslant2$, described by Eq.~\eqref{eq:sk_high}, this contribution vanishes due to the $\cos\phi^{(n)}$ factor, which averages to zero over the azimuthal angle. Therefore, within the approximation considered in Ref.~\cite{Shustin2024en}, the DMI does not influence the total energy of the skyrmion–vortex system. However, deviations from radial symmetry in the HOS, similar to those discussed in subsection~\ref{sec:ansatz_ecc}, may yield a contribution of order $\gamma^2$ and affect the skyrmion’s position $a$. Moreover, Ref.~\cite{%Kuznetsov2023-ru,*
Kuznetsov2023-en} demonstrates that a superconducting film can induce an effective, spatially dispersive DMI in the ferromagnetic layer. Such effective DMI may contribute more significantly to the HOS configurations, even though the “ordinary” DMI remains weak.

\subsection{Stable configurations}

To determine stationary HOS–vortex configurations, one should minimize the total energy. In such a configuration, there exists an optimal value of helicity, determined by the angle $\phi_0$, which depends only on the skyrmion order $n$ and is given by the relation ${\cos(n-1)\phi_0=(-1)^{n}}$. This condition arises from minimizing the first term in Eq.~\eqref{eq:FZ2} for an eccentric configuration (in the coaxial configuration, $\phi_0$ can take arbitrary values due to symmetry).

Next, following Ref.~\cite{Shustin2024en}, we restrict ourselves to an approximate expression for the skyrmion angle in the form of a domain-wall ansatz. This approach may lead to quantitative inaccuracies since it does not account for shape deformation of the skyrmion, but it provides a qualitative description of the variation of the skyrmion radius~$R$, domain wall width~$\delta$, and position~$a$ relative to the vortex.

To analyze the stability of the coaxial configuration, we perform calculations in the limit of small $a\ll\ell_{w}$ and dominant exchange interaction, $A \gg \mathcal{K}_0\sim K$. In this approximation, the total energy can be expressed as:
\begin{equation}
	\mathcal{F}_{\rm tot}\approx \mathcal{F}_0 - \mathcal{F}_1\cdot a^{\alpha_n}+\mathcal{F}_2\cdot a^2,
\end{equation}  
where $\mathcal{F}_{0,1,2}>0$ are constants determined by the system parameters and the optimal values of~$R$ and~$\delta$ at $a=0$. The second term originates from the Zeeman energy. For $n=2$, the exponent $\alpha_{n=2}=1$, and the coaxial state is always unstable, since the free energy decreases with increasing $a$ at small $a$. For $n=3,4$, the exponent is $\alpha_{n=3,4}\approx 2$, so the second and third terms compete. In this case, determining the stable configuration requires a more detailed analysis, partially carried out in~\cite{Shustin2025en}, and also accounting for shape deformations of the skyrmion induced by the vortex magnetic field. In the case $n \geq 5$, formation of coaxial configurations of HOS with a Pearl vortex is expected~\cite{Shustin2025en}, though numerical simulations are necessary to verify this effect.

\section{Discussion\label{Sec:Discuss}}

%----------- SUBSECTION -----------%
\subsection{Experimental studies\label{sec:experiment}}

\begin{table*}[t]
\caption{Experimental data for the exchange stiffness ($A$), magnetic anisotropy constant ($K$), DMI strength ($D$), and saturation magnetization ($M_s$). Based on these parameters, we also compute the characteristic scale $\ell_w$ --- see Eq.~\eqref{eq:epsilon-def}, the dimensionless DMI constant $\epsilon$ --- see Eq.~\eqref{eq:epsilon-def}, and the dimensionless ferromagnetic film parameter $\zeta$ --- see Eq.~\eqref{eq:zeta}.}
    \begin{tabular}{|l|c|c|c|c|c|c|c|c|}
\hline
 & $A$ ($10^{-12}$ J/m) & $K$ ($10^6$ J/m$^3$) & $D$ ($10^{-3}$ J/m$^2$) & $M_s$ ($10^6$ A/m) & $\ell_w $ ($10^{-9}$ m)  & $\epsilon$ & $\zeta$ \\
\hline 
[IrFeCoPt]${}_{10}$~\cite{Petrovic2021} & 13.9 & 1.4 & 2.1 & 1.45 & 3.15 & 0.24 & 8.6 \\
\hline
[Ta/Ir/CoFeB/MgO]${}_7$\cite{Xie2024} & 19 & 0.164 & 1.7 & 1.37 & 11 & 0.48 & 5.8 \\
\hline
Pt/Co/AlO$_x$~\cite{Pizzini2014} & 16 & 0.39 & 2.2 & 1.1 & 6.4 & 0.44 & 5.6 \\
\hline
Ir/Co/Pt~\cite{MoreauLuchaire2016} & $\sim$10 & 0.17 & 1.9 & 0.96 & $\sim$8 & $\sim$0.6 & $\sim$8  \\
\hline
[Co/Pd]${}_n$~\cite{Pollard2017} & 10 & 0.24 & 2.0 & 0.88 & 6 & 0.6 & 7.8\\
\hline
Ta/CoFeB/MgO~\cite{Cao2018} & 10 & 1.3 & 0.22 & 1.2 & 7 & 0.03 & 3.9\\
\hline
Pt/Co/Ni/Co~\cite{Ryu2014} & 20 & 0.6 & 3 & 0.6 & 6 & 0.43 & 2.3\\
\hline
 \end{tabular}
 \label{Tab:exp}
\end{table*}

It turns out that the experimental observation of coupled vortex–skyrmion pairs formed due to stray magnetic fields in a superconductor–chiral magnet heterostructure is a challenging task~\cite{Petrovich2024}. The coexistence of superconducting vortices and Néel-type skyrmions was first observed in a multilayer magnetic structure [Ir$_1$Fe$_{0.5}$Co$_{0.5}$Pt$_1$]$_{10}$/Nb, where the subscript 10 indicates that the heterostructure Ir$_1$Fe$_{0.5}$Co$_{0.5}$Pt$_1$ was repeated 10 times~\cite{Petrovic2021}. A dielectric MgO layer was used to suppress the superconducting proximity effect. Néel-type skyrmions in the structure were stabilized using an external magnetic field and imaged by magnetic force microscopy (MFM). Based on the nonlinear dependence of sample magnetization on the applied magnetic field, the authors concluded that the presence of skyrmions leads to the spontaneous formation of antivortices in the superconductor. Here, an antivortex is defined as a superconducting vortex whose core magnetic field is oriented opposite to the applied external magnetic field. Such spontaneous antivortex generation is consistent with the predictions of Ref.~\cite{Andriyakhina2021} and with the results presented above.

Later, using a similar structure with a different Fe/Co composition, [Ir$_1$Fe$_{0.3}$Co$_{0.7}$Pt$_1$]$_{10}$/NbPt, MFM measurements of Néel skyrmion radii were carried out at two temperatures: above and below the superconducting transition temperature $T_c$~\cite{Machain2021}. It was found that in the superconducting state, the radius of some skyrmions increased by approximately 3\%, which the authors attributed to the spontaneous creation of superconducting antivortices. This increase in skyrmion radius agrees with the predictions of Ref.~\cite{Andriyakhina2021} and with the results discussed in this review.

More recently, a similar MFM experiment at temperatures above and below $T_c$ was conducted on the [Ta/Ir/CoFeB/MgO]$_{7}$/Nb heterostructure~\cite{Xie2024}. The authors discovered that applying a weak magnetic field opposite to the direction of uniform magnetization in the ferromagnet leads, at temperatures below $T_c$, to the emergence of skyrmions with radii nearly twice as large compared to those observed above $T_c$. Conversely, for the opposite field direction, skyrmions of the same radius as above $T_c$ reappeared. The authors explained this behavior by noting that in the first case, a superconducting vortex (referred to as an antivortex in Refs.~\cite{Petrovic2021,Machain2021}) is generated and forms a coaxial configuration with the skyrmion, increasing its radius, while in the second case an antivortex appears and is repelled by the skyrmion. This behavior is in full qualitative agreement with the theoretical results presented in the review. However, for quantitative comparison, it is important to consider that the superconducting film used was relatively thick, and thus the magnetic field of the vortex is not accurately described by the Pearl approximation. Additionally, the multilayer nature of the real heterostructure leads to more complex stray field distributions than those used in our theoretical model.

In summary, Table~\ref{Tab:exp} presents values of various constants for chiral magnets known to support Néel-type skyrmions. Our analysis shows that the characteristic scale $\ell_w$ lies in the range of 5--10~nm, while the dimensionless parameter $\zeta$ is between 2 and 9. A large $\zeta$ allows for efficient tuning of the coupling constant $\gamma = \zeta \ell_w / \lambda = \zeta \ell_w d_S / \lambda_L^2$ over a wide range by varying the thickness $d_S$ of the superconducting film. For example, in experiments~\cite{Petrovic2021,Xie2024} with Nb as the superconductor, the London penetration depth is $\lambda_L \simeq 40$~nm~\cite{Maxfield}, so the Pearl length for a Nb film of thickness $d_S \simeq 10$~nm is $\lambda \simeq 160$~nm. Accordingly, the coupling parameter $\gamma$ for the materials listed in Table~\ref{Tab:exp} will vary in the range of 0.1--0.4.

\subsection{Further directions of theoretical research}

In this review, we focused on the case of a thin superconducting film. However, the developed theory is applicable to any spatial profile of an inhomogeneous magnetic field, including that created by a vortex in a thick superconducting film. In Ref.~\cite{Menezes2019}, the influence of a superconducting antivortex in a thick superconducting film on a skyrmion was studied numerically, and it was found that the skyrmion is stabilized in a coaxial configuration with an increased radius. This indicates that the physical effects discussed above — such as an increase in skyrmion radius and a change in chirality in a coaxial configuration, deformation of the skyrmion in an eccentric configuration, and skyrmion-mediated stabilization of a vortex–antivortex pair — are also possible in thick superconducting films. Naturally, verifying this requires further detailed investigations.

An interesting problem concerns the dynamics of a skyrmion–vortex pair when either the vortex or the skyrmion begins to move under the influence of an applied current in the superconducting or ferromagnetic film, respectively. The first case was numerically studied in Ref.~\cite{Menezes2019} for a skyrmion–antivortex pair. It was found that beyond a critical supercurrent threshold, the pair breaks apart and the antivortex continues to move without the skyrmion. A promising direction would be to determine this threshold current, accounting for possible skyrmion shape deformation during pair motion.

We also note that skyrmion size can change when it moves near various defects, as shown in~\cite{Liu2023} using a simple domain-wall ansatz. However, it would be of interest to apply the variational method developed here, using a more accurate ansatz, to study shape distortions and refine the results obtained in~\cite{Liu2023}.

A powerful tool for probing magnetic states is the study of spin waves. In recent years, significant progress has been made in controlling and manipulating spin waves using external (magnetization gradients~\cite{Nikitov2015}) or internal (domain walls~\cite{Braun1994,Wulfhekel2004,Dijken2018,Laliena2022}, skyrmions~\cite{Nagaosa2014,Garst2014,Aristov2015}) inhomogeneities in the magnetization profile. In a superconductor–ferromagnet heterostructure, the magnon spectrum is influenced by the superconducting film. Several mechanisms underlie this effect: (i) changes in stray fields due to the superconducting film~\cite{Golovchanskiy2018,Golovchanskiy2020,Bauer2022,Silaev2022,Golovchanskiy2023,Borst2023,Kharlan2024}, (ii) the presence of a vortex lattice in the superconductor~\cite{Dobrovolskiy2019,Berakdar2023}, and (iii) the spin-torque transfer mechanism~\cite{Bobkova2022}. It is known~\cite{Garst2014} that skyrmions can support localized magnon states. Similar localized magnon states can also be induced by the magnetic field of a Pearl vortex and its induced distortion of the magnetization profile~\cite{Katkov2024}. It would be interesting to explore which localized magnon states arise in skyrmion–vortex and skyrmion–vortex–antivortex complexes.

As shown in Section~\ref{sec:VaV-Sk}, the presence of a skyrmion affects the interaction between a vortex and an antivortex in a superconducting film. This raises the intriguing question of the interaction between skyrmion and vortex lattices. Recently, Ref.~\cite{Neto2021} addressed this problem via numerical simulations of the equations of motion for superconducting vortices and skyrmions. A rich variety of phases was found. However, in that work, the skyrmions were treated as rigid particles whose parameters do not change in the presence of vortices. It would be interesting to go further and incorporate skyrmion shape deformations during vortex interactions. For lattice systems, this could be done using the stereographic projection approach developed in Refs.~\cite{Aristov2015,Aristov2021,Aristov2022}.

It is well known~\cite{Yablonsky1983en,Mostovoy2006,Dzyaloshinskii2008} that a non-uniform magnetization profile in ferromagnets can lead to a magnetoelectric effect, i.e., the generation of an electric polarization $\bm{P} \propto [(\bm{m}\cdot\nabla)\bm{m}-\bm{m}(\nabla\cdot \bm{m})]$. In an external electric field $\bm{E}$, this results in an additional term in the free energy of the ferromagnetic film, $\int d^2\bm{r}\, \bm{P}\cdot\bm{E}/2$. This opens the possibility to control inhomogeneous magnetic textures via external electric fields. For example, this was demonstrated experimentally in Refs.~\cite{Logginov2008,Pyatakov2015,Pyatakov2015-2}. It would be interesting to apply the theoretical framework developed in this review to explore the effect of external inhomogeneous electric fields — e.g., generated by a charged tip as in the experiments~\cite{Logginov2008,Pyatakov2015,Pyatakov2015-2} — on the shape and chirality of a skyrmion. Given that a skyrmion can be bound to a superconducting vortex, this opens %!% the prospect 
the way %!% 
of indirectly controlling a superconducting vortex via a local electric field.

\section{Conclusion\label{Sec:Concl}}

This review presents the results of a theoretical study on the 
%influence 
effect
of the inhomogeneous magnetic field of a superconducting vortex on a Néel-type skyrmion in thin superconductor–chiral ferromagnet heterostructures. An analytical variational approach is described in detail, based on an ansatz for the skyrmion magnetization profile that accounts for deformations due to the magnetic field of the vortex. This method allowed us to analyze the stability conditions for coaxial vortex–skyrmion configurations and to explain and predict a number of new effects induced by the inhomogeneous magnetic field of the vortex. In the coaxial case, the superconducting vortex can induce a change in the skyrmion's chirality or significantly increase its radius (up to the maximum possible value) while preserving the chirality. For eccentric configurations, the inhomogeneous magnetic field of the vortex leads to a distortion of the cylindrically symmetric skyrmion profile, which is crucial for accurately calculating the equilibrium distance between the centers of the skyrmion and the vortex. The developed approach has also been applied to analyze the stability of a vortex–antivortex–skyrmion triple complex. All theoretical predictions are confirmed by micromagnetic simulations. The review also briefly discusses the effect of the superconducting vortex on high-order skyrmions, recent experiments on the influence of superconducting vortices on Néel skyrmions, and possible directions for future theoretical research.

\vspace{1cm}

{\bf Acknowledgements.} The authors thank P. Vorobyev, D. Katkov, and O. Tretiakov for joint work. The authors are also grateful to A. Kalashnikova, R. Mamin, A. Sadovnikov, Ya. Fominov, A. Fraerman, and M. Shustin for helpful discussions. We thank the Landau Institute for Theoretical Physics for access to its computational cluster, which was used to obtain the numerical simulation results presented here.
This work was supported by the Russian Science Foundation under Grant No. 24-12-00357. It was also partially supported by the Program of Fundamental Research at the National Research University Higher School of Economics.
I. S. Burmistrov %!%acknowledges 
and S. S. Apostoloff acknowledge
personal support from the Foundation for the Advancement of Theoretical Physics and Mathematics “BASIS”.

\bibliography{bib-skyrmion}

\end{document}